\documentclass[twocolumn,tighten]{aastex63}

\usepackage{lineno}

\usepackage{amsmath}

\newcommand{\henir}{{\ion{He}{1}}$~\lambda$10830}

\newcommand{\halpha}{H$\alpha$}
\newcommand{\hbeta}{H$\beta$}
\newcommand{\hgamma}{H$\gamma$}

\usepackage{scalerel}
\newcommand{\CaII}{{\ion{Ca}{2}}}
\newcommand{\CaIIirt}{{\ion{Ca}{2}} IRT}
\newcommand{\CaIIh}{{\ion{Ca}{2}} H}
\newcommand{\CaIIk}{{\ion{Ca}{2}} K}
\newcommand{\CaIItwfi}{{\ion{Ca}{2}}$~\lambda$8498}
\newcommand{\CaIItrfi}{{\ion{Ca}{2}}$~\lambda$8542}

\newcommand{\mdot}{$\dot{\text{M}}$}
\newcommand{\Mdot}{{\dot{{M}}}}

\newcommand{\lsun}{ L_{\sun}}
\newcommand{\msunyr}{M_{\sun} \, \rm{ yr^{-1}}}

\newcommand{\kms}{ \, km \, s^{-1}}
\newcommand{\teff}{T$_{\rm eff}$}

\newcommand{\ri}{R$_{\rm i}$}

\newcommand{\Dr}{$\Delta \rm r$}
\newcommand{\tmax}{T$_{\rm max}$}
\newcommand{\cosi}{$\cos(i)$}

\usepackage{appendix}

\received{December 27, 2022}
\revised{June 12, 2023}
\accepted{June 12, 2023}
\submitjournal{ApJ}

\shorttitle{Ca II lines and disk structure}

\shortauthors{Micolta et al.}

\begin{document}

\title{The \CaII\ lines as 
tracers of disk structure in T Tauri Stars: The Chamaeleon I region}

\correspondingauthor{Marbely Micolta}
\email{micoltam@umich.edu}

\author[0000-0001-8022-4378]{Marbely Micolta}
\affiliation{Department of Astronomy, University of Michigan, 1085 South University Avenue, Ann Arbor, MI 48109, USA}

\author[0000-0002-3950-5386]{Nuria Calvet}
\affiliation{Department of Astronomy, University of Michigan, 1085 South University Avenue, Ann Arbor, MI 48109, USA}

\author[0000-0003-4507-1710]{Thanawuth Thanathibodee}
\affiliation{Institute for Astrophysical Research, Department of Astronomy,
Boston University, 725 Commonwealth Avenue, Boston, MA 02215, US}

\author[0000-0003-1166-5123]{Gladis Magris C.}
\affiliation{Centro de Investigaciones de Astronomía ``Francisco J. Duarte" CIDA, Av. Alberto Carnevali, Mérida 5101, Mérida, Venezuela}

\author[0000-0002-5296-6232]{Mar\'ia Jos\'e Colmenares}
\affiliation{Department of Astronomy, University of Michigan, 1085 South University Avenue, Ann Arbor, MI 48109, USA}
\affiliation{Centro de Investigaciones de Astronomía ``Francisco J. Duarte" CIDA, Av. Alberto Carnevali, Mérida 5101, Mérida, Venezuela}

\author[0000-0002-8237-385X]{Jes\'us V. D\'iaz}
\affiliation{Department of Physics, Western Michigan University, 1903 W. Michigan Avenue, Kalamazoo, MI 49008, USA}
\affiliation{Centro de Investigaciones de Astronomía ``Francisco J. Duarte" CIDA, Av. Alberto Carnevali, Mérida 5101, Mérida, Venezuela}

\author[0000-0002-8351-8854]{Jairo Alzate-Trujillo}
\affiliation{Instituto Nacional de Astrof\'isica, \'Optica y Electr\'onica, Luis Enrique Erro 1, Tonantzintla 72840, Puebla, Mexico.}

\begin{abstract}
We present a study of the \CaIIk\ and IR-triplet lines in a sample of Classical T Tauri stars in the Chamaeleon I star-forming region. We study X-shooter spectra of the stars in the sample and find that in some of these stars the \CaII\ lines are much weaker than expected from their H line fluxes and mass accretion rate. Since the \CaIIk\ lines have characteristic magnetospheric accretion line profiles and the magnetospheric flows feed directly from the inner disk,  we interpret the Ca deficit in terms of depletion due to processes happening in the disk. To test this hypothesis, we define a coarse depletion indicator using the flux of the \CaIIk\ line and show that it correlates with disk properties. In particular, using indicators extracted from Spitzer/IRS spectra, we obtain that all the transitional and pre-transitional disks of the sample show depletion, consistent with trapping of refractories in pressure bumps created by planets and/or in the planets themselves. We find full disks with Ca depletion in the sample that also show indications of advanced dust evolution. We apply magnetospheric accretion models to fit the Balmer and \CaII\ line fluxes of a star showing clear Ca depletion and derive a Ca abundance in its inner disk of about 17\% solar. 
\end{abstract}


\section{Introduction} \label{sec:intro}

Born from molecular clouds, low-mass pre-main sequence stars --- T Tauri Stars or TTS --- are formed surrounded by disks, natural by-products of the star formation process, and nurseries of planets. The disks, left to evolve independently, result in planetary systems. Unraveling the secrets of planet formation, especially in the early stages, has become an outstanding and thrilling goal in astronomy; to achieve this, we need to comprehend in detail the different physical and chemical processes which are at play during the disk evolutionary phases, such as the accretion of matter, material removal through winds or photo-evaporation, dust growth, settling and drifting, dynamical interaction in multiple systems, planet formation, and the interrelation between all these processes.

Each of the mechanisms involved in disk evolution modifies the distribution of materials in the disk, changing the chemical composition of the gas, which
will then be reflected in the material accreting onto the star. For instance, depletion of iron on the stellar surface in Herbig Ae/Be stars (intermediate mass pre-main sequence stars) has been shown to correlate with the trapping of large dust grains in the disks \citep[][]{kama_fingerprints_2015}. In the fully convective structure of low mass TTS, freshly accreted and stellar material rapidly mix, which prevents using stellar abundances to infer inner disk abundance
\citep{jermyn_stellar_2018,kunitomo_revisiting_2018}. Instead, the disk material must be observed directly.

The accretion of the material from the inner region of the protoplanetary disk onto the TTS follows the magnetospheric accretion paradigm \citep[cf.][]{hartmann_accretion_2016}. In this model, the stellar magnetic field truncates the disk and matter is accreted onto the star guided by the field lines at free fall velocities, until it impacts the photosphere in an accretion shock. Typical signatures of accretion in the spectra of T Tauri stars are the excess over the stellar photosphere, specially  at UV wavelengths, due to the accretion shock emission \citep[][]{calvet_structure_1998}, and broad emission lines formed in the accretion flows \citep{muzerolle_emission-line_2001}.
This last property plays an important role in identifying accreting stars and measuring the mass accretion rates 
\citep[][]{hartmann_magnetospheric_1994,muzerolle_brgamma_1998,muzerolle_emission-line_1998, muzerolle_emission-line_2001, white_very_2003,natta_accretion_2004,thanathibodee_magnetospheric_2019}.

Magnetospheric accretion models have been successfully applied to wide ranges of stellar masses, from brown dwarfs \citep{muzerolle_measuring_2005} to Herbig Ae/Be \citep[][]{muzerolle_magnetospheres_2004}, stellar ages, from protostars \citep{muzerolle_brgamma_1998} to 10 Myr old stars \citep{muzerolle_disk_2000},
and levels of accretion, from high accretors \citep{muzerolle_emission-line_2001} to low accretors \citep{thanathibodee_magnetospheric_2019}, providing insight into the physical properties of the accretion process. Magnetospheric accretion links the inner disk to the stellar atmosphere, allowing for a direct determination of chemical abundances in disk material. Abundances in the flows have been inferred from FUV lines 
\citep[][]{herczeg_far-ultraviolet_2002,ardila_hot_2013} and X-ray spectra \citep{kastner_evidence_2002,stelzer_x-ray_2004,drake_x-ray_2005,gunther_x-ray_2006}. The connection between the inner disk and the flows has also been used to probe abundances of volatiles and refractories at the inner edge of the gas disk \citep[][]{mcclure_carbon_2019,mcclure_measuring_2020}. These detailed abundance determinations have been done for only a few bright sources.

Studies of magnetospheric lines so far have focused mostly on the hydrogen lines and little attention has been paid to the Ca II lines, some of the most prominent lines in the spectra of T Tauri stars. In addition to being formed in the magnetospheric accretion flows, the Ca II lines have been found to scale with the accretion rate and to originate in optically thick conditions \citep[e.g.][]{ingleby_accretion_2013,alcala_x-shooter_2014}. Moreover, they provide an exceptional opportunity to probe refractory abundances in the inner disk for large samples of TTS, to complement recent abundance studies  that mainly focus on  volatile species such as CO \citep[][]{ansdell_young_2016,schwarz_radial_2016,oberg_astrochemistry_2021}. By measuring the abundance of refractory elements in the gas falling onto the star, these studies can shed  light on how the evolution of the material in the solid phase proceeds in the disk all the way  to the innermost gas disk.

In this paper, we present the first systematic analysis of the \CaII\ lines for T Tauri stars in the nearby $\sim$2-3 Myr old Chamaeleon I (Cha I) star-forming region \citep[][]{luhman_disk_2008}, a relatively isolated population of young stars with modest levels of extinction \citep[][]{luhman_census_2004}. This sample has the advantage of having observations with the X-shooter spectrograph  \citep[][]{vernet_x-shooter_2011} at the ESO Very Large Telescope (VLT) and also with the  Spitzer Infrared Spectrograph \citep[IRS,][]{houck_infrared_2004}. This makes Cha I an ideal sample for our purposes.

In \S 2 we describe the observational materials and data sources. In \S 3, we analyze the spectroscopic data and derive the characteristics of the \CaII\ lines, and in \S 4 we explore the relationship between Ca abundance and disk structure. In \S 5 we apply magnetospheric accretion models to fit Balmer and \CaII\ line fluxes for the specific case of T28, a star showing a high level of Ca depletion, to get an estimate of the abundance. Finally, in \S 6 we discuss the implications of our results, and in \S 7 we give our conclusions.

\section{Observational Materials} \label{sec:obs}

\subsection{X-shooter sample}

\subsubsection{Classical T Tauri stars} 

We use X-shooter spectra from the Chamaeleon I survey of accreting low mass young stars from \citet[][hereafter M16, M17]{manara_x-shooter_2016, manara_x-shooter_2017}. These studies characterize the stellar properties and determine mass accretion rates (\mdot) by fitting the excess continuum over the photospheric spectrum. This is our sample of accreting T Tauri stars, or Classical T Tauri stars, CTTS. The full M17 sample of CTTS comprises 93 objects and represents 97\% of the disk-bearing stellar objects with spectral types (SpT) earlier than M6 in the region.

We exclude from our sample stars without values of accretion rates or stellar luminosities (L$_*$) in M17, stars with low SNR in the UVB spectra, and stars designated in M17 and M16 as \textit{low accretors}, that is,  objects with UV excess compatible with chromospheric emission, except for T4 which exhibits clear traits of ongoing accretion like red-shifted absorption in the \henir\ line. In addition, due to the compatibility with chromospheric emission, we characterize the objects 2MASS~J11241186-763042 and T51 as low accretors and exclude them from our analysis.

\begin{figure}[!t]
\epsscale{1.2}
\vspace{0.05in}
\plotone{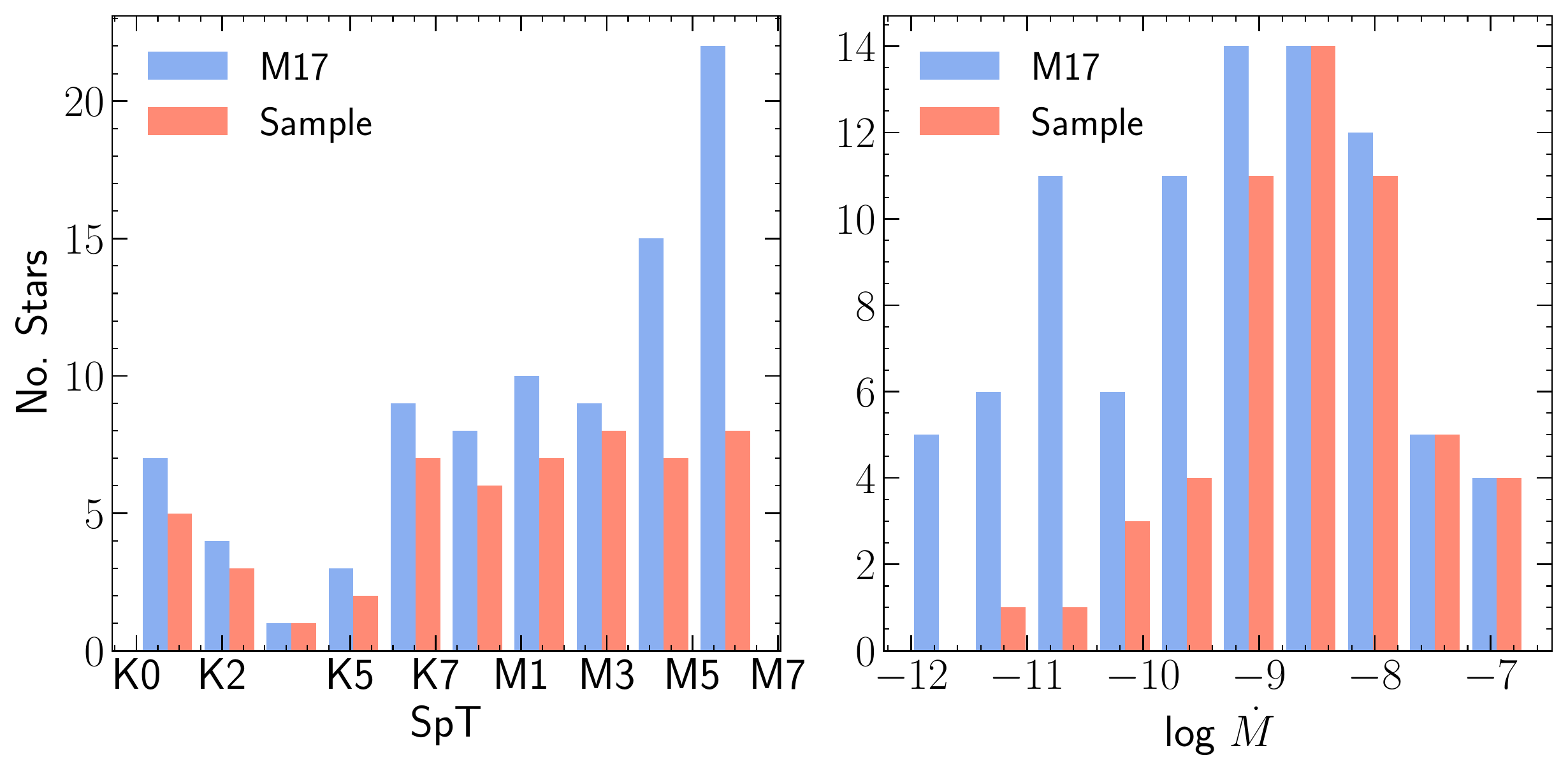}
\caption{Distribution of spectral types and accretion rates for the full Chamaeleon I sample in \citet[][]{manara_x-shooter_2017} (M17) and the sample used in this project.}
\label{hists}
\end{figure}

The final sample consists of 55 stars for which we adopt the stellar parameters from M17, shown in Table \ref{tab:ctts_new}. The distributions of SpT and \mdot\ of our final sample are compared to the M17 sample in Figure \ref{hists}; our selection excludes most of the stars with $\Mdot \leq 10^{-10} \ \msunyr$; however, this does not impact negatively on our results since stars with such low accreting rates show weak lines that require detailed analysis of the profiles to separate the chromospheric and magnetospheric contributions.
We will analyze those in future work.

\begin{deluxetable*}{lcccccccc}[!hbtp]
\tablecaption{Names, stellar parameters and continuum spectral indices for the Cha I sample included in this work
\label{tab:ctts_new}}
\tabletypesize{\scriptsize}

\tablehead{\colhead{Object} & \colhead{2MASS} & \colhead{SpT$^1$} & \colhead{Teff$^1$} & \colhead{Av$^1$} & \colhead{log \mdot$^1$} & \colhead{n$_{2-6}$$^2$} & \colhead{n$_{13-31}$$^2$} & \colhead{Disk Type$^{1,2}$}}
\startdata
... & J11065939-7530559 & M5.5 & 3060 & 0.4 & -11.12 & -1.75 & 0.37 & ... \\
... & J11085367-7521359 & M1 & 3705 & 1.5 & -8.15 & ... & ... & ... \\
... & J11183572-7935548 & M5 & 3125 & 0.0 & -8.95 & ... & ... & TD \\
... & J11432669-7804454 & M5.5 & 3060 & 0.4 & -8.71 & ... & ... & ... \\
CHX18N & J11114632-7620092 & K2 & 4900 & 0.8 & -8.09 & ... & ... & ... \\
CHXR 47 & J11103801-7732399 & K4 & 4590 & 3.9 & -8.12 & -1.87 & -0.66 & ... \\
CR Cha & J10590699-7701404 & K0 & 5110 & 1.3 & -8.71 & -1.82 & -0.12 & ... \\
CS Cha & J11022491-7733357 & K2 & 4900 & 0.8 & -8.29 & -2.62 & 2.9 & TD \\
CT Cha A & J11040909-7627193 & K5 & 4350 & 2.4 & -6.69 & -1.15 & -0.32 & ... \\
CW Cha & J11123092-7644241 & M0.5 & 3780 & 2.1 & -8.03 & ... & ... & ... \\
Cha-Ha-2 & J11074245-7733593 & M5.5 & 3060 & 2.4 & -10.05 & -1.66 & -0.99 & ... \\
Cha-Ha-6 & J11083952-7734166 & M6.5 & 2935 & 0.1 & -10.25 & ... & ... & ... \\
ESO-Ha-562 & J11080297-7738425 & M1 & 3705 & 3.4 & -9.24 & ... & ... & ... \\
Hn 10e & J11094621-7634463 & M3 & 3415 & 2.1 & -9.51 & -1.41 & 0.37 & ... \\
Hn 5 & J11064180-7635489 & M5 & 3125 & 0.0 & -9.28 & -0.67 & -1.44 & ... \\
Hn13 & J11105597-7645325 & M6.5 & 2935 & 1.3 & -9.57 & -1.58 & -0.4 & ... \\
Hn21W & J11142454-7733062 & M4.5 & 3200 & 2.2 & -9.04 & -1.98 & -0.77 & ... \\
ISO-ChaI-143 & J11082238-7730277 & M5.5 & 3060 & 1.3 & -10.07 & -1.5 & -1.04 & ... \\
ISO-ChaI-282 & J11120351-7726009 & M5.5 & 3060 & 2.8 & -9.89 & -1.16 & -0.54 & ... \\
Sz Cha & J10581677-7717170 & K2 & 4900 & 1.3 & -7.82 & -2.07 & 1.72 & TD/PTD \\
Sz18 & J11071915-7603048 & M2 & 3560 & 1.3 & -8.7 & -2.75 & 2.8 & TD \\
Sz19 & J11072074-7738073 & K0 & 5110 & 1.5 & -7.63 & -1.35 & -0.67 & ... \\
Sz22 & J11075792-7738449 & K5 & 4350 & 3.2 & -8.34 & -1.06 & -0.32 & ... \\
Sz27 & J11083905-7716042 & K7 & 4060 & 2.9 & -8.86 & -2.01 & 1.47 & TD/PTD \\
Sz32 & J11095340-7634255 & K7 & 4060 & 4.3 & -7.08 & -0.73 & 0.0 & ... \\
Sz33 & J11095407-7629253 & M1 & 3705 & 1.8 & -9.35 & -1.36 & -0.37 & ... \\
Sz37 & J11104959-7717517 & M2 & 3560 & 2.7 & -7.82 & -1.46 & -0.27 & ... \\
Sz45 & J11173700-7704381 & M0.5 & 3780 & 0.7 & -8.09 & -2.13 & 0.9 & TD/PTD \\
T10 & J11004022-7619280 & M4 & 3270 & 1.1 & -9.22 & ... & ... & ... \\
T12 & J11025504-7721508 & M4.5 & 3200 & 0.8 & -8.7 & ... & ... & ... \\
T16 & J11045701-7715569 & M3 & 3415 & 4.9 & -7.8 & ... & ... & ... \\
T23 & J11065906-7718535 & M4.5 & 3200 & 1.7 & -8.11 & -1.7 & -0.92 & ... \\
T24 & J11071206-7632232 & M0 & 3850 & 1.5 & -8.49 & -1.76 & -0.22 & ... \\
T27 & J11072825-7652118 & M3 & 3415 & 1.2 & -8.36 & -1.74 & -0.78 & ... \\
T28 & J11074366-7739411 & M1 & 3705 & 2.8 & -7.92 & -1.61 & -0.6 & ... \\
T3 & J10555973-7724399 & K7 & 4060 & 2.6 & -8.61 & -0.81 & -0.49 & ... \\
T3 B & ... & M3 & 3415 & 1.3 & -8.43 & ... & ... & ... \\
T30 & J11075809-7742413 & M3 & 3415 & 3.8 & -8.31 & ... & ... & ... \\
T33 B & J11081509-7733531 & K0 & 5110 & 2.7 & -8.69 & ... & ... & ... \\
T37 & J11085090-7625135 & M5.5 & 3060 & 0.8 & -10.74 & ... & ... & ... \\
T38 & J11085464-7702129 & M0.5 & 3780 & 1.9 & -9.3 & -1.35 & -0.19 & ... \\
T4 & J10563044-7711393 & K7 & 4060 & 0.5 & -9.41 & ... & ... & ... \\
T40 & J11092379-7623207 & M0.5 & 3780 & 1.2 & -7.33 & -1.41 & -0.91 & ... \\
T44 & J11100010-7634578 & K0 & 5110 & 4.1 & -6.68 & -0.9 & 0.42 & ... \\
T45 & J11095873-7737088 & M0.5 & 3780 & 3.0 & -6.95 & -1.15 & -1.02 & ... \\
T46 & J11100704-7629376 & K7 & 4060 & 1.2 & -8.7 & -1.53 & -1.16 & ... \\
T48 & J11105333-7634319 & M3 & 3415 & 1.2 & -7.96 & -1.29 & -0.75 & ... \\
T49 & J11113965-7620152 & M3.5 & 3340 & 1.0 & -7.41 & -1.48 & -0.38 & ... \\
T5 & J10574219-7659356 & M3 & 3415 & 1.4 & -8.51 & -1.73 & -0.32 & ... \\
T50 & J11120984-7634366 & M5 & 3125 & 0.1 & -9.34 & -1.85 & -0.33 & ... \\
T51 & J11122441-7637064 & K2 & 4900 & 0.1 & -8.16 & -1.09 & -1.52 & ... \\
T51 B & ... & M2 & 3560 & 0.5 & -9.07 & ... & ... & ... \\
T52 & J11122772-7644223 & K0 & 5110 & 1.0 & -7.48 & -1.41 & -0.27 & ... \\
TW Cha & J10590108-7722407 & K7 & 4060 & 0.8 & -8.86 & -1.56 & -0.17 & ... \\
VW Cha & J11080148-7742288 & K7 & 4060 & 1.9 & -7.6 & -1.48 & -0.17 & ...
\enddata
\tablecomments{Parameters obtained from 1: \cite{manara_x-shooter_2017} and 2: \cite{manoj_spitzer_2011}, typical errors for Teff are $\pm 70\ K$ for M-type stars and $\pm 145\ K$ for earlier stars. (TD) refers to Transitional Disks. (PTD) refers to Pre-Transitional disks.}  (...) refers to full disks.
\end{deluxetable*}

\subsubsection{Weak-line T Tauri stars}

T Tauri stars are magnetically very active, which implies that their chromospheres may contribute to the flux in lines and continua. Therefore, the chromospheric contribution needs to be taken into consideration in the analysis of their spectra. To account for this, we use the library of photospheric templates of pre-main sequence, non-accreting stars (Weak-line T Tauri stars, WTTS) in \citet[][]{manara_x-shooter_2013,manara_extensive_2017}. We restricted the sample to stars within the spectral type range of our sample of CTTS. The objects RX J0457.5+2014, TWA6, TWA13A, Sz122, Sz121, and Sz107 were excluded because of their large \halpha\ width at 10\% height. Analogously to the CTTS, we adopt the stellar parameters from  the literature (Table \ref{tab:wtts_t1}).

\begin{deluxetable*}{lccccccccc}[!htb]
\tablecaption{Names and stellar parameters for the WTTS sample included in this work \label{tab:wtts_t1}}
\tabletypesize{\scriptsize}
\tablehead{\colhead{Object} & \colhead{SpT} & \colhead{Av} & \colhead{d (pc)} & \colhead{T$_{\mathrm{eff}}$} & \colhead{$v \sin{i}$} & \colhead{$\sigma$ $v \sin{i}$} & \colhead{Ref.}}
\startdata
HBC 407 & K0 & 0.8 & 140 & 5110 & 10.0 & 1.0 & 1 \\
LM 717 & M6.5 & 0.4 & 160 & 2935 & 19.0 & 21.0 & 1 \\
PZ99 J160550.5-253313 & K1 & 0.7 & 145 & 5000 & 13.0 & 6.0 & 1 \\
PZ99 J160843.4-260216 & K0.5 & 0.7 & 145 & 5050 & 42.5 & 1.1 & 1 \\
Par-Lup3-2 & M5 & 0.00 & 200 & 3125 & 26.4 & 4.7 & 2 \\
RX J0438.6+1546 & K2 & 0.20 & 140 & 4900 & 26.3 & 1.0 & 1 \\
RX J1515.8-3331 & K0.5 & 0.00 & 150 & 5050 & 22.3 & 1.0 & 1 \\
RX J1538.6-3916 & K4 & 0.40 & 150 & 4590 & 1.0 & 2.0 & 1 \\
RX J1540.7-3756 & K6 & 0.10 & 150 & 4205 & 19.1 & 1.0 & 1 \\
RX J1543.1-3920 & K6 & 0.10 & 150 & 4205 & 12.1 & 1.0 & 1 \\
RX J1547.7-4018 & K3 & 0.10 & 150 & 4730 & 11.1 & 1.0 & 1 \\
SO641 & M5 & 0.38 & 360 & 3125 & 12.5 & 4.9 & 2 \\
SO797 & M4.5 & 0.14 & 360 & 3200 & 27.4 & 0.9 & 2 \\
SO879 & K7 & 0.28 & 360 & 4060 & 10.0 & 4.0 & 2 \\
SO925 & M5.5 & 0.00 & 360 & 3060 & 20.8 & 4.2 & 2 \\
SO999 & M5.5 & 0.00 & 360 & 3060 & 116.3 & 4.0 & 2 \\
Sz94 & M4 & 0.19 & 200 & 3270 & 29.8 & 2.7 & 2 \\
TWA13B & M1 & 0.19 & 59 & 3705 & 10.4 & 1.1 & 2 \\
TWA14 & M0.5 & 0.00 & 96 & 3780 & 46.0 & 0.7 & 2 \\
TWA15B & M2 & 0.00 & 111 & 3415 & 21.6 & 0.9 & 2 \\
TWA25 & M0 & 0.00 & 54 & 3850 & 14.4 & 0.5 & 2 \\
TWA2A & M2 & 0.28 & 47 & 3560 & 18.0 & 1.9 & 2 \\
TWA7 & M2 & 0.00 & 28 & 3415 & 8.4 & 1.3 & 2 \\
TWA9A & K5 & 0.09 & 68 & 4350 & 14.2 & 0.4 & 2 \\
TWA9B & M3 & 0.19 & 68 & 3415 & 11.6 & 2.2 & 2 \\
TWA15A & M3.5 & 0.00 & 111 & 3340 & 33.6 & 0.9 & 2
\enddata
\tablecomments{Parameters adopted from 1: \citet[][]{manara_extensive_2017} and 2: \citet[][]{manara_x-shooter_2013}, typical errors for Teff are $\pm 70\ K$ for M-type stars and $\pm 145\ K$ for earlier stars.} For the stars in \citet[][]{manara_x-shooter_2013}, the extinction was obtained from the photometry available in (2) and the $v \sin{i}$ values were adopted from \citet[][]{stelzer_x-shooter_2013}

\end{deluxetable*}

\subsection{Spitzer sample}

We cross-matched our Cha I spectroscopic sample with the mid-IR sample from \citet[][]{manoj_spitzer_2011}, which presents a detailed analysis of the 
5-30 $\mu$m {\it Spitzer IRS} spectra of 62 Class II objects in the region, finding 42 stars in common. We adopted their results for the continuum spectral indices n$_{2-6}$ and  n$_{13-31}$ to probe disk evolution in the planet-formation region of our CTTS sample; the indices are reported in Table \ref{tab:ctts_new}. Additionally, we adopt their classification for TW Cha, CR Cha, and T52 as stars with enhanced 10 $\mu$m silicate emission, based on their measurements of the equivalent width of the feature.

\section{Data Analysis} \label{sec: data}

Here we present the analysis of the X-shooter sample. All the spectra were corrected for reddening using the standard reddening law $\rm R_v = 3.1$ \citep{cardelli_relationship_1989}. We excluded the {\CaIIh} line from this analysis, since it can be partially or fully blended with the H$\epsilon$ line.

\subsection{Line profiles} \label{sec: profiles}

\begin{figure*}[t!]
\epsscale{0.85}
\plotone{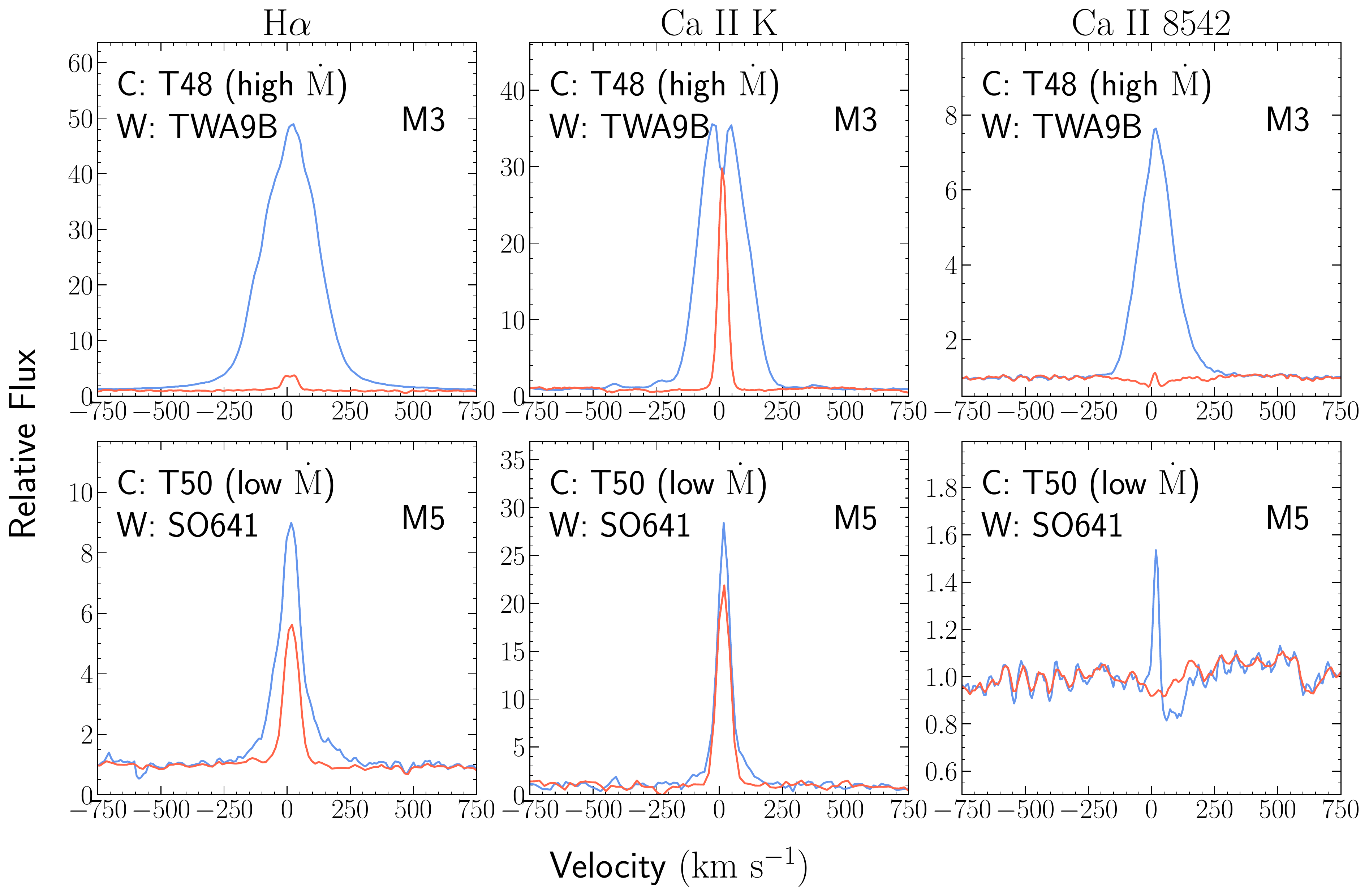}
\caption{Comparison between \halpha, \CaIIk, and \CaIItrfi, one of the IR triplet lines, for
CTTS (blue) and WTTS (red) of the same SpT. The top panel shows the high accretor T48 ($log\ \Mdot = -7.96$) and the WTTS TWA9B. The bottom panel shows low accretor T50 ($log\ \Mdot = -9.07$) and the WTTS SO641.}
\label{profiles}
\end{figure*}

Figure \ref{profiles} shows a comparison of the profiles of {\halpha}, {\CaIIk}, and {\CaIItwfi}, one of the Ca II IR triplet (IRT) lines, in a CTTS and a WTTS of the same spectral type. The top panel shows profiles of the high accretor T48 ($log\ \Mdot = -7.96$), while the bottom panel shows profiles of the low accretor T50 ($log\ \Mdot = -9.07$); the profiles of the WTTS TWA9B and SO641, respectively, are also included for comparison. Note that we show fluxes relative to the continuum, which is higher than the photospheric continuum, specially at short wavelengths for the accreting stars. Therefore, the main comparison here is between line profiles and not total relative fluxes.

We can see from Figure \ref{profiles} that the \CaII\ lines behave similarly 
to \halpha\ for high \mdot, showing broad wings, as expected \citep{white_very_2003}. In contrast, for low \mdot, \halpha\ still has high-velocity wings, though weaker than in high \mdot\ case, but the \CaII\ lines are essentially chromospheric; their narrow profiles are comparable to those of WTTS, yet they still show traits of magnetospheric accretion; e.g., in Figure \ref{profiles} T50 shows red-shifted absorption in the \CaIItrfi\ line and a trace of high velocity wings in \CaIIk\ line.

\begin{figure*}[!t]
\epsscale{0.85}
\plotone{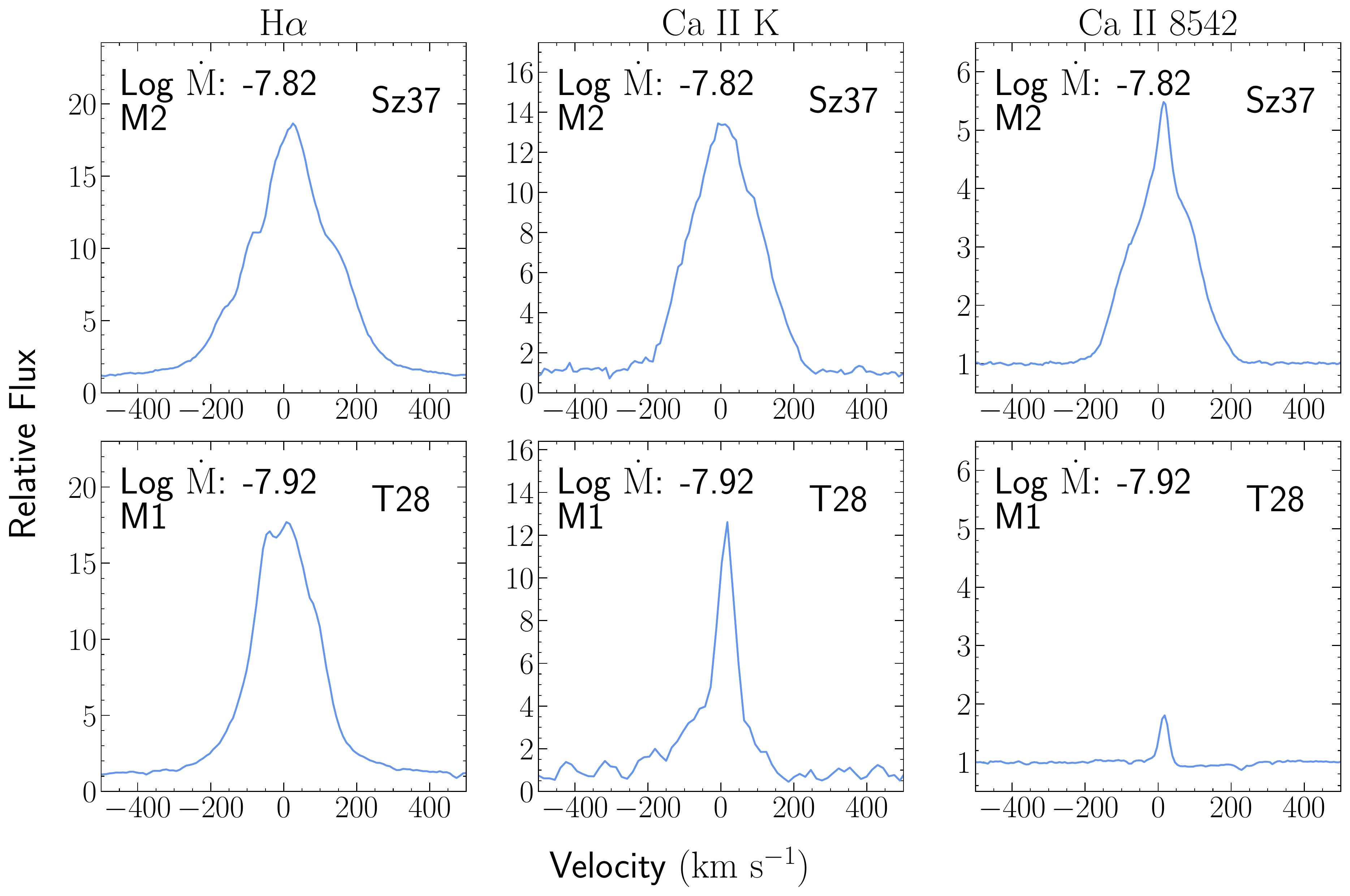}
\caption{Comparison between \halpha, \CaIIk, and \CaIItrfi, one of the IR triplet lines, for two stars in Cha I with similar physical conditions i.e., similar stellar parameters. Accretion rates in units of $\msunyr$. T28 is a Ca-poor star (see text).}
\label{ca-poor profile}
\end{figure*}

Differing from the expected behavior, we found cases where high accretors, with consistently strong and broad \halpha\ profiles, had weak, chromospheric-like \CaII\ lines. An example of these \textbf{Ca-poor} cases is shown in Figure \ref{ca-poor profile}, in which the \halpha, \CaIIk, and \CaIItwfi\ profiles of the stars T28 and Sz37 are compared. Within the uncertainties, both stars have similar stellar parameters and mass accretion rates, determined from the UV flux excess over the photospheric fluxes \citep{manara_extensive_2017}. They also have comparable fluxes in the H$\alpha$ line, as expected from the similarities of the accretion rates. The differences in the shape of the profiles may be due to inclination, which is unknown for these stars, and/or the geometry of the magnetosphere. The similarities of {\mdot} and \halpha\ flux suggest that the stars have comparable densities and temperatures in the magnetospheric flows and accretion shocks, in which the \halpha\ line and the excess UV emission originate \citep{muzerolle_emission-line_2001,calvet_structure_1998}.

In contrast to Sz37, T28 shows much narrower and weak \CaII\ lines; in fact, they are essentially chromospheric as one would expect from a low accretor. Only the \CaIIk\ line shows a hint of a magnetospheric component, namely a high-velocity “pedestal” in addition to a chromospheric peak, similar to the \halpha\ profile of very low accretors \citep{thanathibodee_census_2023}. Since the densities and temperatures of the magnetospheric flows of T28 are high, as shown by its \halpha\ profile and its mass accretion rate, the observed weakness of the \CaII\ lines strongly suggests an absence of Ca, i.e., \emph{Ca depletion}, in the magnetospheric flows, and therefore in the inner gas disk of T28. On this basis, we characterize T28 as Ca-poor, and hereafter we will use it as our standard for a star with \ion{Ca}{0} depletion.

If this is the case, the depletion would be due to processes in the disk, such as pressure bumps or planet formation \citep[e.g.][]{kama_fingerprints_2015}, that may have sequestered refractory material further out from the dust wall. To test this possibility, we explore if there is a relation between disk properties and the depletion we observe in the spectra in \S \ref{depletion}.

\subsection{Line Fluxes and Luminosities}

We calculated line fluxes by integrating extinction-corrected, continuum-subtracted line profiles. Prior to the calculations, the contribution of the photosphere to the \CaIIirt\ was removed for all the stars, along with the photospheric component for the \CaIIk\ and Hydrogen lines for the early-K stars. For this purpose, we compared the observed spectra with CIFIST2011\_2015 synthetic BT-Settl spectra \citep[][]{baraffe_new_2015,allard_atmospheres_2012} of effective temperature (\teff) within 50 K of the star and a typical value of log $g = 4.0$, convolved at the same resolution of X-shooter and rotationally broadened at the same rotational velocity ($v\sin i$) as the object.

To estimate $v\sin i$, we used the Fourier method \citep{carroll_spectroscopic_1933}, which requires unblended photospheric lines at a sufficient signal-to-noise ratio (SNR) to have reliable line shapes \citep{simon-diaz_fourier_2007}.

We selected 9 photospheric lines in the spectrum, derived the $v\sin i$ for each one, and adopted the mean value and the standard deviation as the final measurement and its error (Table \ref{tab:ctts_hvsini}). 
We also calculated the  $v\sin i$ including  \ion{Li}{1}\,$\lambda6708$ as a 10th line; the differences in the values obtained by including this line are less than the error for all stars but two, T3 B and T33 B, for which the difference is still below 10\%. 
In four stars of our sample, the photospheric lines could not be successfully detected, i.e., for all lines the line center depth was less than $3\sigma$. In the case of 2MASSJ11065939-7530559 and ISO-ChaI-138, we attribute this to low SNR of the spectrum. In in the case of Sz 22 and T44, no lines were detected due to high veiling diluting the lines. In these cases, the Fourier method was no longer applicable, and we adopted the average value of $v\sin i = 15 \kms$ \citep[e.g.][]{hartmann_additional_1989, covino_study_1997}, verifying the validity of this approximation when convolving and comparing with the BT-Settl models.

We compared our results for $v\sin i$ with the ones provided in \citet{frasca_gaia-ESO_2015} for 26 stars in common. We note our method results in higher  $v\sin i$ values, with the differences being smaller for earlier spectral types and/or lower accretion rates. In particular, for the early K stars, our results fall within 5\% of the ones in 
\citet{frasca_gaia-ESO_2015}. We also note that the differences in the resulting line fluxes fall within the typical error of 20\%.

We followed the method used in \citet[][]{alcala_x-shooter_2014} to determine the line flux, calculating three independent measurements per line, 
corresponding to the lowest, highest, and middle position of the local continuum depending on the local noise level of the spectra. Subsequently, the flux and its error were computed as the average and standard deviation of the three independent measurements, respectively. The extinction-corrected fluxes and their errors are provided in Tables \ref{tab:ctts_hvsini}, \ref{tab:ctts_ca}, and \ref{tab:wtts_t2}. The luminosity of the emission lines follows as $L_{\rm line} = 4\pi d^2 \ F_{\rm line}$ where $d$ is the distance to the star. We use $d = 160$pc \citep[][]{luhman_disk_2008} 
for consistency between this work and the M17 analysis, from which we adopted the stellar parameters.

\begin{deluxetable*}{lcccccccccc}[!hbtp]
\tablecaption{Hydrogen lines fluxes and $v \sin{i}$ derived for the Cha I sample included in this work \label{tab:ctts_hvsini}}
\tabletypesize{\scriptsize}
\tablehead{\colhead{Object} & \colhead{$v \sin{i}$} & \colhead{$\sigma$ $v \sin{i}$} & \colhead{F(H$\alpha$)} & \colhead{$\sigma$(H$\alpha$)} & \colhead{F(H$\beta$)} & \colhead{$\sigma$(H$\beta$)} & \colhead{F(H$\gamma$)} & \colhead{$\sigma$(H$\gamma$)}}
\startdata
CHX18N & 29.0 & 1.9 & 172.0 & 19.5 & 21.5 & 9.3 & 9.1 & 4.1 \\
CHXR 47 & 24.6 & 2.5 & 77.6 & 34.0 & 9.7 & 10.9 & 7.0 & 6.4 \\
CR Cha & 36.1 & 1.0 & 1140.0 & 70.9 & 137.0 & 44.3 & 48.0 & 24.8 \\
CS Cha & 19.6 & 4.9 & 709.0 & 34.3 & 80.3 & 19.7 & 53.1 & 11.1 \\
CT Cha A & 17.7 & 5.7 & 1390.0 & 72.0 & 472.0 & 58.4 & 237.0 & 32.9 \\
CW Cha & 26.7 & 6.0 & 237.0 & 6.4 & 52.1 & 4.8 & 36.2 & 4.6 \\
Cha-Ha-2 & 27.3 & 0.7 & 6.3 & 0.1 & 1.2 & 0.0 & 1.0 & 0.0 \\
Cha-Ha-6 & 21.2 & 4.5 & 2.0 & 0.0 & 0.2 & 0.0 & 0.1 & 0.0 \\
ESO-Ha-562 & 24.2 & 1.3 & 89.6 & 1.7 & 17.0 & 0.4 & 9.2 & 0.4 \\
Hn 10e & 17.8 & 6.6 & 17.8 & 0.4 & 4.4 & 0.1 & 3.3 & 0.1 \\
Hn 5 & 19.8 & 4.8 & 5.2 & 0.2 & 2.4 & 0.1 & 2.2 & 0.1 \\
Hn13 & 18.5 & 5.5 & 2.6 & 0.1 & 0.3 & 0.0 & 0.2 & 0.0 \\
Hn21W & 19.2 & 4.7 & 10.5 & 0.5 & 1.4 & 0.1 & 0.9 & 0.1 \\
ISO-ChaI-143 & 26.9 & 2.6 & 4.4 & 0.1 & 0.5 & 0.0 & 0.5 & 0.0 \\
ISO-ChaI-282 & 27.5 & 0.9 & 2.8 & 0.1 & 0.8 & 0.0 & 0.7 & 0.0 \\
J11065939-7530559 & 15.0 & 0.0 & 0.7 & 0.0 & 0.2 & 0.0 & 0.2 & 0.0 \\
J11085367-7521359 & 17.3 & 5.7 & 176.0 & 4.8 & 73.3 & 2.8 & 68.9 & 2.6 \\
J11183572-7935548 & 38.1 & 5.4 & 10.9 & 0.5 & 2.4 & 0.1 & 1.3 & 0.1 \\
J11432669-7804454 & 21.6 & 2.6 & 45.3 & 0.8 & 6.2 & 0.2 & 4.1 & 0.2 \\
Sz Cha & 36.3 & 3.4 & 301.0 & 33.0 & 17.1 & 11.8 & 8.1 & 6.0 \\
Sz18 & 18.0 & 5.1 & 45.8 & 2.9 & 4.3 & 0.8 & 3.3 & 0.5 \\
Sz19 & 33.7 & 3.7 & 1180.0 & 105.0 & 91.1 & 48.1 & 6.6 & 8.5 \\
Sz22 & 15.0 & 0.0 & 273.0 & 8.9 & 42.6 & 4.2 & 16.9 & 2.6 \\
Sz27 & 20.5 & 4.6 & 191.0 & 6.1 & 10.7 & 2.2 & 7.6 & 1.3 \\
Sz32 & 38.7 & 0.6 & 2350.0 & 39.8 & 682.0 & 33.1 & 352.0 & 32.8 \\
Sz33 & 19.5 & 4.6 & 25.5 & 1.4 & 5.4 & 0.6 & 4.4 & 0.4 \\
Sz37 & 17.6 & 2.5 & 274.0 & 6.5 & 100.0 & 4.5 & 61.9 & 4.4 \\
Sz45 & 17.9 & 5.6 & 272.0 & 6.4 & 51.6 & 3.3 & 37.3 & 2.5 \\
T10 & 14.2 & 7.3 & 41.7 & 0.9 & 6.4 & 0.3 & 4.8 & 0.2 \\
T12 & 19.8 & 5.5 & 33.6 & 0.8 & 8.4 & 0.2 & 7.3 & 0.2 \\
T16 & 26.7 & 0.9 & 27.0 & 3.4 & 20.0 & 1.9 & 27.7 & 1.8 \\
T23 & 18.8 & 5.8 & 140.0 & 2.3 & 25.6 & 0.7 & 17.3 & 0.6 \\
T24 & 16.7 & 5.9 & 38.6 & 6.1 & 9.4 & 2.6 & 8.2 & 1.8 \\
T27 & 22.7 & 6.5 & 68.1 & 3.3 & 17.5 & 1.0 & 14.4 & 0.8 \\
T28 & 16.3 & 5.2 & 312.0 & 7.6 & 50.6 & 3.2 & 40.4 & 2.4 \\
T3 & 26.7 & 0.9 & 137.0 & 4.1 & 28.4 & 2.6 & 25.0 & 2.1 \\
T3 B & 23.3 & 4.1 & 32.9 & 1.8 & 14.6 & 0.9 & 12.4 & 0.7 \\
T30 & 25.6 & 2.1 & 126.0 & 2.7 & 22.3 & 1.0 & 16.6 & 0.9 \\
T33 B & 35.5 & 1.0 & 276.0 & 15.9 & 24.9 & 9.6 & 5.0 & 3.7 \\
T37 & 20.8 & 4.7 & 1.9 & 0.1 & 0.4 & 0.0 & 0.3 & 0.0 \\
T38 & 25.5 & 3.5 & 39.0 & 2.0 & 8.3 & 0.8 & 5.7 & 0.5 \\
T4 & 21.0 & 5.7 & 25.0 & 3.4 & 8.0 & 1.6 & 6.3 & 0.9 \\
T40 & 23.8 & 4.2 & 471.0 & 15.7 & 172.0 & 11.9 & 127.0 & 11.3 \\
T44 & 15.0 & 0.0 & 6490.0 & 166.0 & 1830.0 & 122.0 & 956.0 & 93.4 \\
T45 & 22.1 & 6.0 & 1240.0 & 26.5 & 407.0 & 20.6 & 331.0 & 22.7 \\
T46 & 14.3 & 5.6 & 206.0 & 9.5 & 19.6 & 3.4 & 14.9 & 2.5 \\
T48 & 24.6 & 2.8 & 396.0 & 3.4 & 99.6 & 1.5 & 58.9 & 1.3 \\
T49 & 18.9 & 6.0 & 237.0 & 5.3 & 53.7 & 4.9 & 40.5 & 4.9 \\
T5 & 19.1 & 5.3 & 20.0 & 2.0 & 3.7 & 0.5 & 2.6 & 0.3 \\
T50 & 14.5 & 7.1 & 7.0 & 0.4 & 1.9 & 0.1 & 1.4 & 0.0 \\
T51 B & 19.3 & 5.2 & 19.2 & 1.0 & 7.4 & 0.5 & 6.7 & 0.4 \\
T52 & 26.0 & 2.1 & 3010.0 & 101.0 & 772.0 & 68.0 & 356.0 & 40.2 \\
TW Cha & 17.9 & 1.4 & 122.0 & 5.2 & 35.0 & 2.3 & 26.9 & 1.5 \\
VW Cha & 26.9 & 2.0 & 1130.0 & 32.0 & 165.0 & 14.3 & 105.0 & 10.1
\enddata
\tablecomments{Fluxes in units of $10^{-14} \rm erg \ s^{-1} \ cm^{-2}$. Velocities in units of 
${\rm Km \, s^{-1}}$}
\end{deluxetable*}
\begin{deluxetable*}{lcccccccc}[!hbtp]
\tablecaption{Calcium II lines fluxes derived for the Cha I sample included in this work \label{tab:ctts_ca}}
\tabletypesize{\scriptsize}
\tablehead{\colhead{Object} & \colhead{F($\lambda 3933$)} & \colhead{$\sigma$($\lambda 3933$)} & \colhead{F($\lambda 8498$)} & \colhead{$\sigma$($\lambda 8498$)} & \colhead{F($\lambda 8542$)} & \colhead{$\sigma$($\lambda 8542$)} & \colhead{F($\lambda 8662$)} & \colhead{$\sigma$($\lambda 8662$)}}
\startdata
CHX18N & 19.8 & 4.0 & 9.8 & 3.4 & 11.7 & 3.0 & 11.4 & 3.3 \\
CHXR 47 & 43.4 & 10.5 & 23.5 & 11.8 & 25.7 & 9.9 & 20.4 & 8.0 \\
CR Cha & 94.6 & 25.8 & 27.1 & 13.8 & 33.8 & 12.2 & 25.0 & 9.7 \\
CS Cha & 52.5 & 10.2 & 14.3 & 4.6 & 17.2 & 4.3 & 16.3 & 4.6 \\
CT Cha A & 3150.0 & 136.0 & 874.0 & 24.9 & 906.0 & 20.1 & 728.0 & 19.6 \\
CW Cha & 43.3 & 5.2 & 15.7 & 1.9 & 18.8 & 1.5 & 17.0 & 1.5 \\
Cha-Ha-2 & 0.3 & 0.0 & 0.4 & 0.2 & 0.4 & 0.1 & 0.4 & 0.2 \\
Cha-Ha-6 & 0.0 & 0.0 & 0.1 & 0.0 & 0.1 & 0.0 & 0.1 & 0.0 \\
ESO-Ha-562 & 13.6 & 0.3 & 6.9 & 0.7 & 7.4 & 0.5 & 6.8 & 0.6 \\
Hn 10e & 2.2 & 0.1 & 1.0 & 0.2 & 1.3 & 0.2 & 1.0 & 0.2 \\
Hn 5 & 1.5 & 0.1 & 0.8 & 0.2 & 0.8 & 0.1 & 0.7 & 0.2 \\
Hn13 & 0.2 & 0.0 & 0.5 & 0.1 & 0.4 & 0.1 & 0.3 & 0.1 \\
Hn21W & 0.8 & 0.0 & 0.9 & 0.4 & 0.6 & 0.3 & 0.5 & 0.3 \\
ISO-ChaI-143 & 0.3 & 0.0 & 0.3 & 0.1 & 0.3 & 0.1 & 0.3 & 0.1 \\
ISO-ChaI-282 & 0.5 & 0.0 & 0.4 & 0.1 & 0.3 & 0.1 & 0.3 & 0.1 \\
J11065939-7530559 & 0.1 & 0.0 & 0.1 & 0.0 & 0.1 & 0.0 & 0.1 & 0.0 \\
J11085367-7521359 & 20.1 & 3.2 & 4.7 & 1.4 & 6.6 & 1.1 & 6.0 & 1.2 \\
J11183572-7935548 & 1.2 & 0.1 & 1.5 & 0.7 & 1.0 & 0.5 & 0.9 & 0.6 \\
J11432669-7804454 & 2.4 & 0.1 & 1.6 & 0.5 & 1.9 & 0.4 & 1.8 & 0.5 \\
Sz Cha & 26.0 & 11.8 & 10.9 & 5.9 & 13.8 & 5.4 & 13.0 & 4.8 \\
Sz18 & 3.0 & 0.2 & 2.1 & 0.9 & 1.9 & 0.5 & 2.1 & 0.7 \\
Sz19 & 144.0 & 78.4 & 41.6 & 23.3 & 54.0 & 17.7 & 41.9 & 14.1 \\
Sz22 & 75.7 & 5.5 & 50.9 & 3.2 & 54.6 & 2.6 & 46.2 & 2.5 \\
Sz27 & 4.6 & 0.5 & 4.2 & 1.4 & 5.7 & 1.3 & 5.0 & 1.2 \\
Sz32 & 1200.0 & 79.6 & 716.0 & 11.9 & 711.0 & 10.3 & 587.0 & 10.3 \\
Sz33 & 2.1 & 0.2 & 1.6 & 0.6 & 2.1 & 0.5 & 1.9 & 0.5 \\
Sz37 & 117.0 & 5.6 & 52.6 & 2.3 & 58.1 & 1.8 & 49.4 & 1.9 \\
Sz45 & 16.2 & 2.1 & 7.7 & 2.3 & 10.3 & 1.7 & 8.5 & 1.7 \\
T10 & 1.3 & 0.1 & 0.8 & 0.2 & 0.8 & 0.2 & 0.7 & 0.2 \\
T12 & 1.8 & 0.1 & 1.2 & 0.4 & 1.0 & 0.3 & 0.9 & 0.4 \\
T16 & 18.4 & 1.7 & 5.0 & 1.3 & 5.0 & 1.0 & 4.7 & 1.1 \\
T23 & 29.7 & 0.6 & 19.0 & 1.2 & 21.5 & 1.0 & 18.5 & 1.0 \\
T24 & 5.7 & 0.4 & 3.7 & 1.7 & 4.2 & 1.5 & 4.0 & 1.6 \\
T27 & 10.3 & 0.7 & 3.7 & 1.6 & 3.8 & 1.4 & 3.0 & 1.2 \\
T28 & 17.0 & 1.2 & 8.5 & 2.8 & 11.7 & 2.3 & 11.0 & 2.4 \\
T3 & 15.1 & 1.7 & 6.9 & 1.8 & 10.7 & 1.6 & 9.5 & 1.6 \\
T3 B & 6.5 & 0.6 & 2.4 & 0.7 & 2.5 & 0.5 & 2.0 & 0.6 \\
T30 & 27.7 & 0.9 & 11.8 & 1.1 & 13.6 & 0.9 & 12.5 & 1.0 \\
T33 B & 47.1 & 11.6 & 19.8 & 5.4 & 22.1 & 4.5 & 17.0 & 3.8 \\
T37 & 0.1 & 0.0 & 0.2 & 0.1 & 0.2 & 0.1 & 0.2 & 0.1 \\
T38 & 3.6 & 0.3 & 2.9 & 0.8 & 3.7 & 0.8 & 3.2 & 0.7 \\
T4 & 5.3 & 0.2 & 5.4 & 2.0 & 6.9 & 1.9 & 6.1 & 1.8 \\
T40 & 83.5 & 13.3 & 62.6 & 4.3 & 71.8 & 3.6 & 61.5 & 3.3 \\
T44 & 4360.0 & 233.0 & 1740.0 & 46.5 & 1820.0 & 39.7 & 1480.0 & 36.0 \\
T45 & 360.0 & 25.1 & 90.1 & 6.3 & 112.0 & 5.3 & 88.6 & 4.9 \\
T46 & 5.5 & 0.7 & 5.7 & 2.7 & 7.8 & 2.5 & 7.7 & 2.4 \\
T48 & 91.0 & 1.9 & 38.1 & 1.3 & 43.6 & 1.0 & 37.3 & 1.2 \\
T49 & 9.3 & 4.7 & 5.0 & 1.5 & 6.1 & 1.0 & 5.3 & 1.1 \\
T5 & 3.4 & 0.1 & 2.7 & 1.0 & 2.3 & 0.7 & 1.8 & 0.7 \\
T50 & 0.9 & 0.0 & 1.3 & 0.3 & 0.9 & 0.2 & 0.7 & 0.2 \\
T51 B & 1.5 & 0.2 & 1.5 & 0.4 & 1.8 & 0.4 & 1.7 & 0.4 \\
T52 & 919.0 & 57.1 & 496.0 & 30.0 & 603.0 & 25.3 & 469.0 & 22.8 \\
TW Cha & 15.9 & 1.0 & 10.4 & 2.5 & 14.4 & 2.3 & 12.5 & 2.2 \\
VW Cha & 59.0 & 4.3 & 42.1 & 11.6 & 60.5 & 10.3 & 51.8 & 9.6
\enddata
\tablecomments{Fluxes in units of $10^{-14} \rm erg \ s^{-1} \ cm^{-2}$.}

\end{deluxetable*}
\begin{longrotatetable}
\begin{deluxetable*}{lcccccccccccccc}
\tablecaption{Fluxes derived for the WTTS sample included in this work \label{tab:wtts_t2}}
\tabletypesize{\scriptsize}
\tablehead{\colhead{Object} & \colhead{F(H$\alpha$)} & \colhead{$\sigma$(H$\alpha$)} & \colhead{F(H$\beta$)} & \colhead{$\sigma$(H$\beta$)} & \colhead{F(H$\gamma$)} & \colhead{$\sigma$(H$\gamma$)} & \colhead{F($\lambda 3933$)} & \colhead{$\sigma$($\lambda 3933$)} & \colhead{F($\lambda 8498$)} & \colhead{$\sigma$($\lambda 8498$)} & \colhead{F($\lambda 8542$)} & \colhead{$\sigma$($\lambda 8542$)} & \colhead{F($\lambda 8662$)} & \colhead{$\sigma$($\lambda 8662$)}}
\startdata
HBC 407 & 24.1 & 11.5 & -16.8 & 8.9 & -24.2 & 6.27 & 29.2 & 14.9 & 10.5 & 47.8 & 13.9 & 48.2 & 10.0 & 47.1 \\
LM 717 & 1.4 & 0.03 & 0.2 & 3E-3 & 0.1 & 8E-4 & 0.08 & 4E-3 & 0.2 & 0.7 & 0.2 & 0.8 & 0.2 & 1.1 \\
PZ99 J160550.5-253313 & 156.0 & 47.0 & 18.8 & 40.6 & -8.87 & 29.7 & 260.0 & 69.4 & 60.3 & 180.0 & 71.2 & 173.0 & 72.3 & 173.0 \\
PZ99 J160843.4-260216 & 221.0 & 65.7 & -99.9 & 61.5 & -139.0 & 48.5 & 456.0 & 123.0 & 97.2 & 245.0 & 130.0 & 235.0 & 121.0 & 236.0 \\
Par-Lup3-2 & 10.7 & 0.4 & 2.4 & 0.07 & 1.6 & 0.04 & 1.3 & 0.08 & 1.4 & 6.04 & 1.7 & 6.2 & 1.7 & 7.6 \\
RX J0438.6+1546 & 186.0 & 32.9 & 30.8 & 21.9 & -0.4 & 13.6 & 151.0 & 25.7 & 65.0 & 147.0 & 77.1 & 141.0 & 82.9 & 146.0 \\
RX J1515.8-3331 & 114.0 & 33.2 & -11.5 & 23.5 & -29.7 & 16.1 & 135.0 & 36.4 & 46.9 & 145.0 & 61.9 & 143.0 & 61.7 & 144.0 \\
RX J1538.6-3916 & 82.1 & 20.3 & 4.7 & 13.7 & 2.9 & 9.2 & 108.0 & 16.9 & 32.9 & 87.1 & 44.2 & 83.8 & 47.6 & 86.1 \\
RX J1540.7-3756 & 59.8 & 9.4 & 20.7 & 4.5 & 12.9 & 2.7 & 54.9 & 3.8 & 30.5 & 49.3 & 38.4 & 47.5 & 33.5 & 49.9 \\
RX J1543.1-3920 & 71.1 & 8.9 & 34.1 & 4.6 & 25.4 & 2.8 & 58.0 & 3.9 & 33.4 & 47.5 & 41.2 & 45.9 & 36.9 & 47.9 \\
RX J1547.7-4018 & 70.0 & 23.2 & -4.4 & 15.8 & -9.1 & 10.4 & 109.0 & 20.0 & 39.6 & 104.0 & 51.1 & 99.9 & 55.8 & 103.0 \\
SO641 & 1.2 & 0.03 & 0.3 & 4E-3 & 0.2 & 2E-3 & 0.2 & 4E-3 & 0.2 & 0.4 & 0.2 & 0.4 & 0.07 & 0.5 \\
SO797 & 1.8 & 0.05 & 0.4 & 8E-3 & 0.2 & 4E-3 & 0.3 & 7E-3 & 0.3 & 0.8 & 0.4 & 0.8 & 0.2 & 0.9 \\
SO879 & 24.5 & 1.9 & 8.1 & 0.8 & 4.1 & 0.4 & 13.0 & 0.5 & 10.3 & 11.0 & 12.8 & 10.8 & 11.2 & 11.3 \\
SO925 & 0.8 & 0.02 & 0.1 & 2E-3 & 0.05 & 9E-4 & 0.09 & 2E-3 & 0.1 & 0.3 & 0.2 & 0.3 & 0.12 & 0.4 \\
SO999 & 2.03 & 0.04 & 0.3 & 5E-3 & 0.2 & 2E-3 & 0.2 & 5E-3 & 0.25 & 0.7 & 0.3 & 0.7 & 0.1 & 0.9 \\
Sz94 & 27.3 & 0.8 & 9.1 & 0.2 & 5.9 & 0.1 & 7.1 & 0.2 & 3.8 & 9.0 & 3.3 & 9.1 & 3.2 & 10.0 \\
TWA13B & 302.0 & 20.3 & 117.0 & 6.8 & 73.2 & 3.6 & 144.0 & 5.07 & 101.0 & 156.0 & 110.0 & 153.0 & 96.9 & 161.0 \\
TWA14 & 157.0 & 4.8 & 39.1 & 1.45 & 22.8 & 0.74 & 37.9 & 1.07 & 31.3 & 34.2 & 40.1 & 34.8 & 34.4 & 36.6 \\
TWA15B & 69.9 & 1.4 & 20.9 & 0.3 & 11.3 & 0.2 & 15.0 & 0.3 & 11.7 & 14.4 & 12.7 & 14.7 & 10.3 & 16.2 \\
TWA25 & 528.0 & 31.0 & 215.0 & 11.9 & 133.0 & 6.3 & 224.0 & 8.4 & 150.0 & 201.0 & 169.0 & 200.0 & 160.0 & 200.0 \\
TWA2A & 570.0 & 41.1 & 197.0 & 12.5 & 118.0 & 6.9 & 279.0 & 10.3 & 199.0 & 342.0 & 192.0 & 348.0 & 203.0 & 365.0 \\
TWA7 & 476.0 & 17.5 & 132.0 & 3.8 & 74.9 & 1.9 & 136.0 & 3.36 & 128.0 & 190.0 & 118.0 & 193.0 & 88.0 & 211.0 \\
TWA9A & 205.0 & 18.1 & 78.9 & 9.0 & 46.9 & 5.3 & 125.0 & 7.31 & 78.3 & 108.0 & 106.0 & 108.0 & 100.0 & 110.0 \\
TWA9B & 72.5 & 2.6 & 24.7 & 0.7 & 15.6 & 0.4 & 21.5 & 0.6 & 16.3 & 27.8 & 17.2 & 28.4 & 13.0 & 31.2 \\
TWA15A & 103.0 & 1.5 & 31.2 & 0.4 & 18.7 & 0.2 & 18.8 & 0.3 & 13.8 & 15.7 & 13.6 & 16.0 & 9.8 & 17.8
\enddata
\tablecomments{Fluxes in units of $10^{-15} \rm erg \ s^{-1} \ cm^{-2}$}.

\end{deluxetable*}
\end{longrotatetable}

\subsubsection{Correlations with accretion rates} \label{sec:corr}

\begin{figure}[!t]
\epsscale{1.2}
\vspace{0.05in}
\plotone{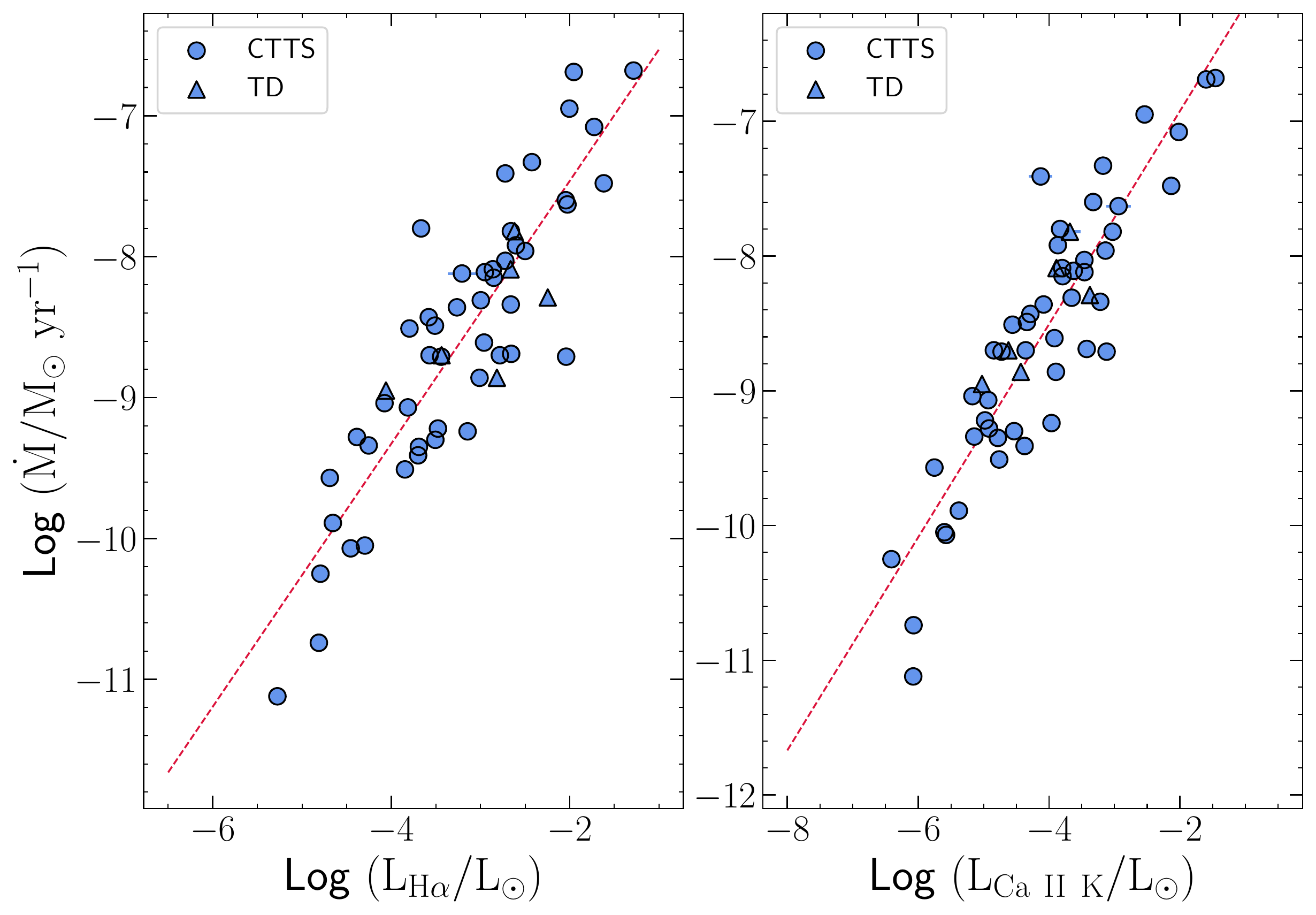}
\caption{\mdot\ vs. \halpha\ and \CaIIk\ line luminosities. Fits to the data are given in eq. \ref{mdot-halpha} and \ref{mdot-cak} respectively.}

\label{stats}
\end{figure}

\begin{figure*}[!t]
\epsscale{0.9}
\vspace{0.05in}
\plotone{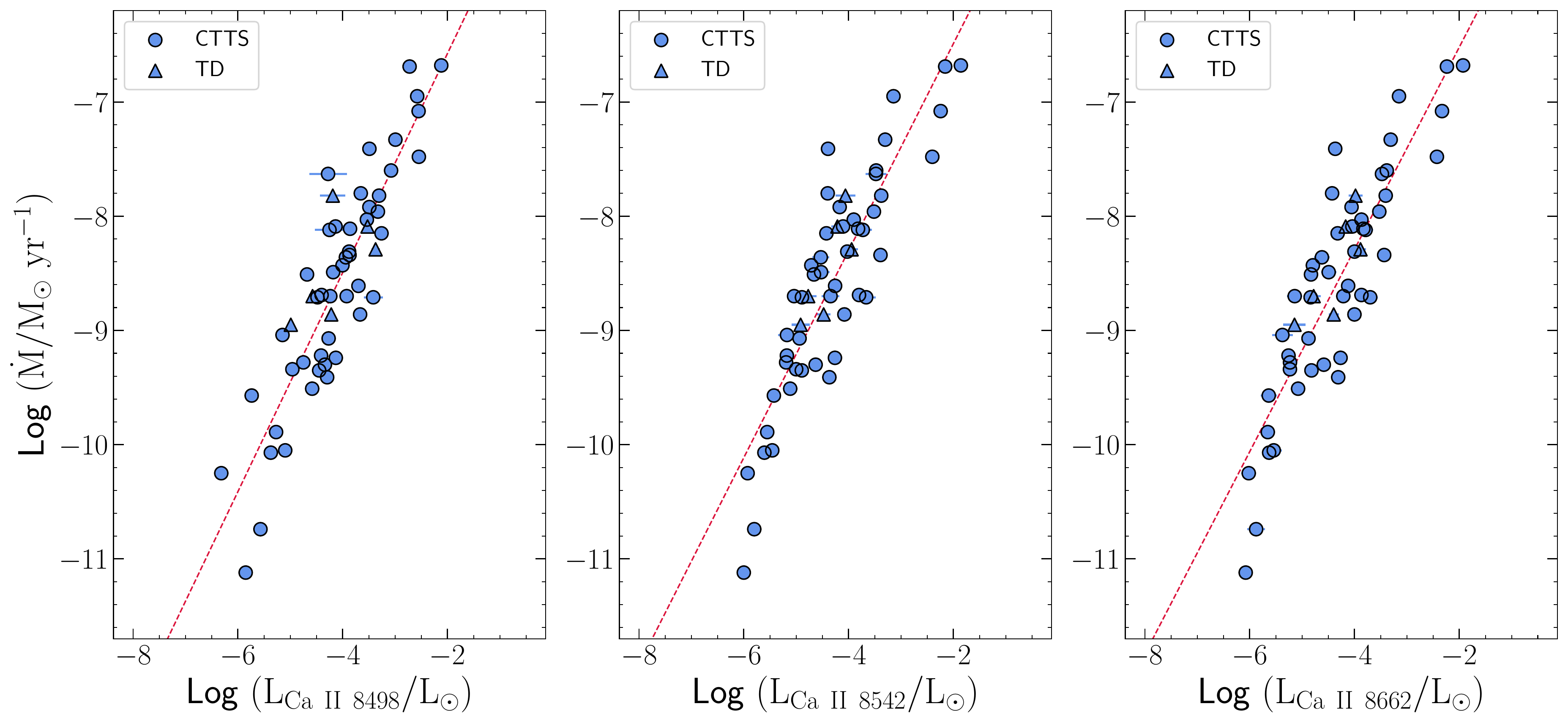}
\caption{\mdot\ vs. \CaIIirt\ line luminosities. Fits to the data are given in eq. \ref{mdot-ca25}, \ref{mdot-ca35} and \ref{mdot-ca24}.}
\label{stats-irt}
\end{figure*}

Previous works have found correlations between line luminosities and \mdot\ or the accretion luminosity (L$_{acc} $) \citep[e.g.,][]{ingleby_accretion_2013,alcala_x-shooter_2017}. Here, we expand the analysis to our Cha I sample by computing linear least-squares regressions.
We firstly conduct the analysis for \halpha, usually accepted as a good tracer of the accretion in stars, obtaining a Pearson correlation coefficient of 0.87 between the \halpha\ luminosity and \mdot; the fit to the data is shown with a dashed line in the left panel of Figure \ref{stats} and is described by the following equation:

\vspace{-0.05in}

\begin{equation}
        \log \left( \frac{\dot{M}}{\msunyr}\right) = 
        0.93\ (\pm 0.07)\ \log \left( \frac{L_{H\alpha}}{\lsun} \right)  -\ 5.6\ (\pm 0.2)
        \label{mdot-halpha}
\end{equation}

\vspace{0.1in}

Similarly, we explore the correlations between \CaII\ luminosities and {\mdot}.
The right panel of Figure \ref{stats} shows the correlation between the \CaIIk\ line and {\mdot}, finding a Pearson correlation coefficient of 0.90. The  least-squares fit is shown in the right panel of Figure \ref{stats} and is given by:

\vspace{-0.05in}
\begin{equation} \label{mdot-cak}
   \log \left( \frac{\dot{M}}{\msunyr}\right) = 0.79\ (\pm 0.05)\ \log \left( \frac{L_{C\MakeLowercase{a}\protect\scaleto{II}{1.2ex}K} }{ \lsun} \right)  - 5.3\ (\pm 0.2)
\end{equation}

Our trends are consistent with those of \citet[][]{ingleby_accretion_2013} within the uncertainties.

Analogously, we inspect the correlations  between \mdot\ and the \CaIIirt\ for our sample, shown in Figure \ref{stats-irt}. We obtain a 0.88 
Pearson coefficient for the three cases, with a least-squares
fit given by the following relationships: 

\begin{equation}  \label{mdot-ca25}
     \log \left( \frac{\dot{M}}{\msunyr}\right) = 0.94\ (\pm 0.07)\ \log \left( \frac{ L_{C\MakeLowercase{a} \lambda 8498} }{ \lsun} \right) - 4.7\ (\pm 0.3)
\end{equation}

\vspace{-0.05in}

\begin{equation}  \label{mdot-ca35}
    \log \left( \frac{\dot{M}}{\msunyr}\right) = 0.91\ (\pm 0.07)\ \log \left( \frac{ L_{C\MakeLowercase{a} \lambda 8542} }{ \lsun} \right) - 4.7\ (\pm 0.3)
\end{equation}

\vspace{-0.05in}

\begin{equation}  \label{mdot-ca24}
    \log \left( \frac{\dot{M}}{\msunyr}\right) = 0.89\ (\pm 0.07)\ \log \left( \frac{ L_{C\MakeLowercase{a} \lambda 8662} }{ \lsun} \right) - 4.7\ (\pm 0.3)
\end{equation}

\vspace{0.1in}

Overall, the ratio between the \CaIIirt\ lines is very close to 1:1:1, consistent with optically thick gas conditions, while Balmer decrements \halpha/\hbeta\ have a median of 4.8 with maximum values $\sim\ 10$, more consistent with optically thin gases. The wide range of physical conditions in which the emission lines in T Tauri stars arise has already been noted in previous work \citep{herbig_observations_1980,hamann_emission_1992,alcala_x-shooter_2014,frasca_gaia-ESO_2015,frasca_x-shooter_2017}, and they have been compared to the differences between plages and prominences in the Sun. This diversity favors the magnetospheric accretion model over uniform and isothermal slabs, given the wide range of temperatures and densities in the accretion flows, which include those found in solar active regions \citep[cf.][]{muzerolle_emission-line_2001}.

Moreover, we see a deviation from the main trend at low accretion rates, exhibiting line luminosities higher than expected for the \mdot. This behavior is possibly caused by the chromospheric contribution in the lines, which we will explore in the following section.

\subsubsection{Chromospheric component in CTTS \label{chro}}

\begin{figure}[!t]
\epsscale{1.2}
\vspace{0.05in}
\plotone{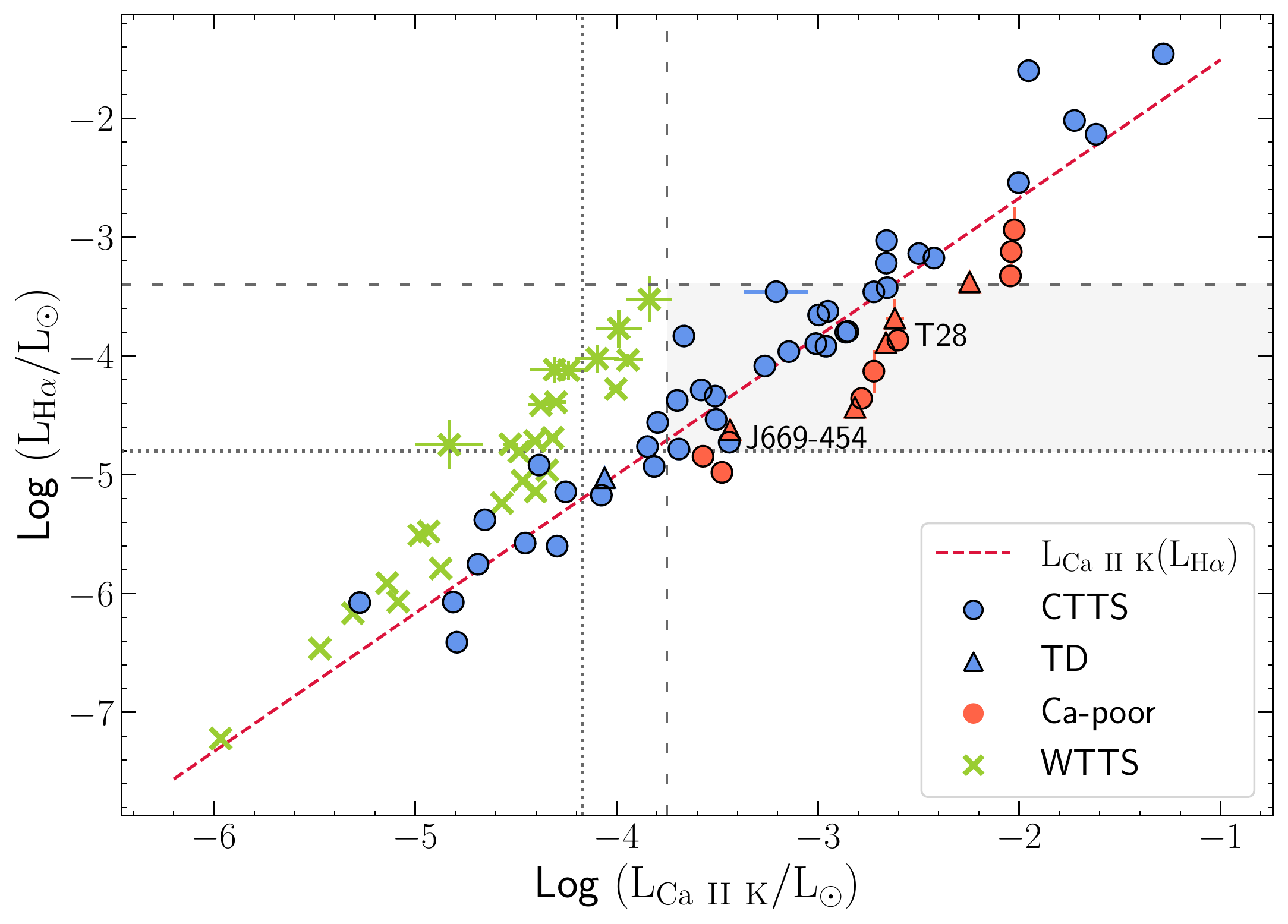}
\caption{Comparison of L\CaIIk\ vs. \halpha\ for CTTS (blue) and WTTS (green). We show in red the CTTS identified as Ca-poor stars (see text).  Dotted and dashed gray lines mark the thresholds where CTTS and WTTS have comparable luminosities and the maximum contribution observed of the chromosphere for K stars or later, respectively. The dashed red line represents the general CTTS trend and is given by eq. (\ref{halpha-cak}).
}
\label{ctts-wtts}
\end{figure}

In this section, we explore the increasing relevance of the chromospheric emission for low accretors mentioned in sections \ref{sec: profiles} and \ref{sec:corr}. Figure \ref{ctts-wtts} shows a comparison between the luminosities of \halpha\ and \CaIIk\ in CTTS and WTTS, where the dashed line is given by the equation:

\vspace{-0.05in}
\begin{equation}
     \log \left( \frac{L_{C\MakeLowercase{a}\protect\scaleto{II}{1.2ex}K} }{ \lsun} \right) = 1.2\ (\pm 0.2)\ \log \left( \frac{L_{H\alpha}}{\lsun} \right) - 0.3\ (\pm 0.6)
    \label{halpha-cak}
\end{equation}

Obtained by matching equations \ref{mdot-halpha} and \ref{mdot-cak}.

As expected, CTTS reach higher line luminosities than WTTS due to the contribution from the magnetospheric accretion. However, for low  line luminosities,
and correspondingly, lower accretion rates, the \CaIIk\ line luminosities of the
CTTS become comparable to those of the WTTS, which indicates that the line emission is becoming increasingly dominated by the chromosphere, c.f. Figure \ref{profiles}.

We find that we cannot distinguish the CTTS from WTTS if the star has $\log\ \left( {L_{H\alpha}}/{\lsun} \right) \leq -4.17$ and  $\log\ \left( {L_{Ca II K}}/{\lsun} \right) \leq -4.8$, shown as dotted gray lines in Figure \ref{ctts-wtts}. Below these thresholds, the CTTS analysis based on flux is no longer applicable; instead, a very detailed analysis of the profiles is required as the chromosphere is the dominant component of the line (c.f. Bottom row in Figure \ref{profiles}). Above these limits, accreting stars and non-accreting stars follow different trends. We find that the maximum contribution of the chromosphere for a CTTS of SpT K or later is $\log\ \left( {L_{H\alpha}}/{\lsun} \right) \sim  -3.75$ and  $\log\ \left( {L_{Ca II K}}/{\lsun} \right) \sim -3.4$, shown as gray dashed lines in Figure \ref{ctts-wtts}.

We find a similar behavior for the \CaII\ IRT lines; in general, the \CaII\ lines are significantly more sensitive to chromospheric emission than  the \halpha\ line, especially the \CaIIirt\ lines, becoming relevant at higher \mdot\ values. This is expected because the opacity of the \CaII\ lines is smaller than those of the hydrogen lines, making the emission highly dependent on the gas density; therefore, compared to the hydrogen lines, they need higher accretion rates (i.e., higher densities) for the magnetospheric component to fully dominate the emission line luminosity. We highlight this behavior as a gray region in Figure \ref{ctts-wtts}. The CTTS inside this zone show {\halpha} luminosities above the maximum value for chromospheric contribution (dashed gray line) yet their {\CaIIk} luminosities are still comparable with the ones of the chromospheres of K stars.

On this basis, we focus on CTTS with {\halpha} above the maximum value for chromospheric contribution (dashed gray line), so we can guarantee their accretion status, to identify possible Ca-poor stars. For this purpose, we inspect their line profiles, as we did in \S \ref{sec: profiles}, searching for stars on which the Ca II profiles looked more of a chromospheric origin than in stars with similar spectral types and accretion rates. With this restriction, in addition to T28, we identify stars CR Cha, CS Cha, 
Sz Cha, Sz18, Sz19, Sz27, Sz45, T10, T12, T46, T49, VW Cha, as Ca-poor and show them in red in Figure \ref{ctts-wtts}. Their profiles are shown in Figures \ref{fig-a: prof 1}, \ref{fig-a: prof 2}, and \ref{fig-a: prof 3}. These stars sit further down from the expected trend (red dashed line), reflecting that they show less \CaIIk\ flux than expected from their mass accretion rates. 

We note the star J11432669-7804454 (J668-454 in  Figure \ref{ctts-wtts}) falls within Ca-poor stars. However, its line profiles do not show obvious signs of Ca depletion, with \CaII\ line profiles having a significant contribution of magnetospheric accretion at the base, and only showing the chromospheric narrow component at the core of the line (see Figure \ref{fig-a: prof 4}). This may be an intermediate abundance case, but detailed line profile modeling is needed to discern the origins of this behavior; we will address it in future work.

\section{C\lowercase{a} depletion} \label{depletion}

In previous sections, we showed that the Ca II lines could be much weaker relative to the H lines in CTTS of similar mass accretion rates. We interpreted this effect as due to Ca depletion in the magnetospheric flows, where these lines arise, and identified  the stars with the most evident case of depletion. Here, we aim to quantify the depletion and explore the origin of this phenomenon, making use of the disk properties known for the Cha I sample. For this purpose, hereafter we focus only on the CTTS with \halpha\ luminosities mostly from magnetospheric accretion, i.e., $\log\ \left( {L_{H\alpha}}/{\lsun} \right) >  -3.75$, restricting our final sample to 40 stars.

\subsection{A depletion indicator \label{dep_index}}

\begin{figure}[!t]
\epsscale{1.2}
\vspace{0.05in}
\plotone{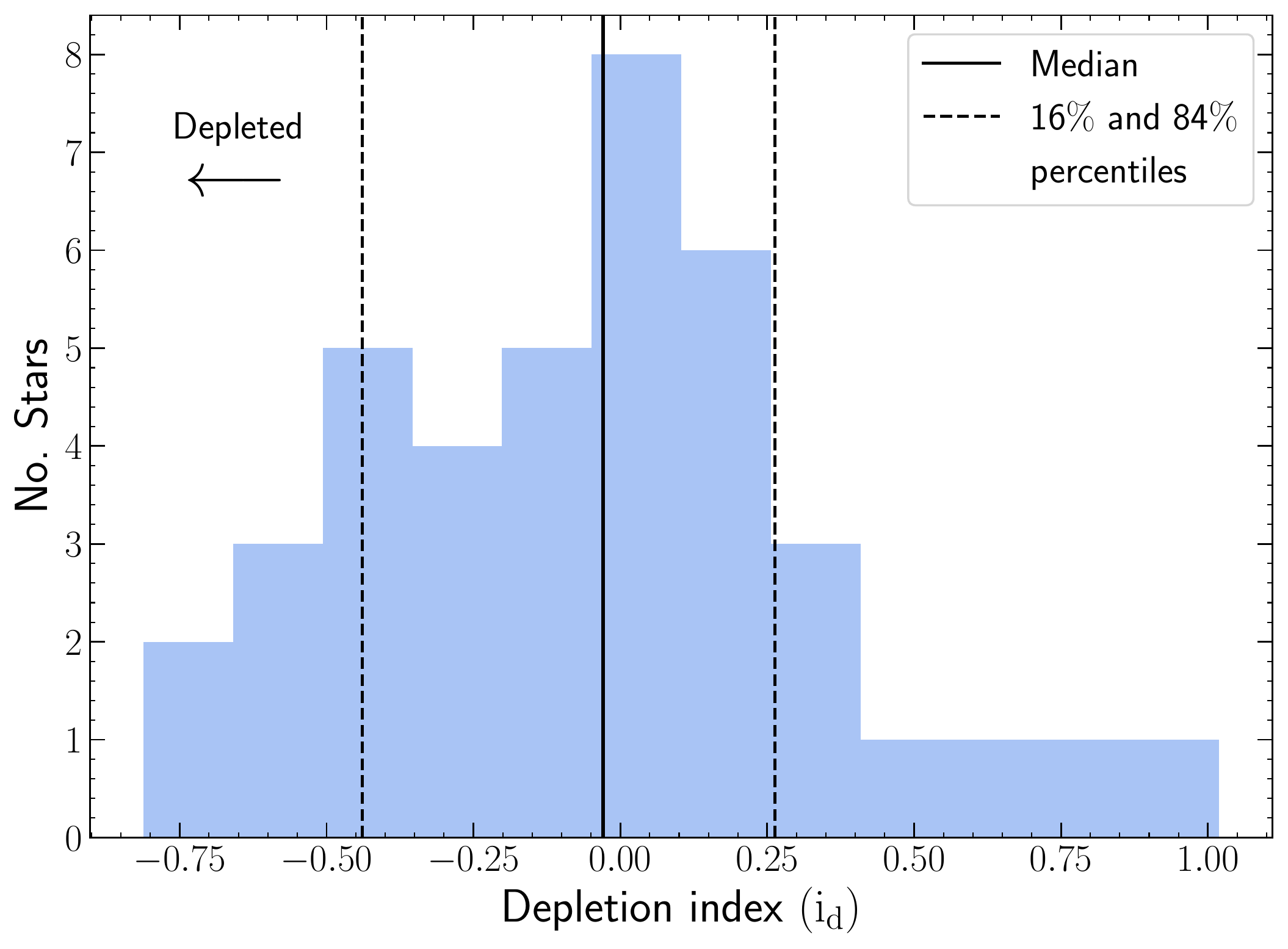}
\caption{Distribution of the depletion index for the Cha I sample. The black dashed line indicates the median. Black dashed lines mark 16\% and 84\% percentiles.}
\label{dep_hist}
\end{figure}

To quantify the depletion seen in the spectra we focus on the behavior of the \CaIIk\ line, since this line is strong enough in our selected sample to show a magnetospheric component, even with a high depletion factor (c.f. Figure \ref{ca-poor profile}). To obtain an estimate of how much Ca depletion a star shows, we define a depletion index given by:

\begin{equation}
    i_d = \left( \frac{L_{C\MakeLowercase{a}\protect\scaleto{II}{1.2ex}K}^{obs} }{L_{C\MakeLowercase{a}\protect\scaleto{II}{1.2ex}K}^{*} } \right) 
\end{equation}

Where $L_{C\MakeLowercase{a}\protect\scaleto{II}{1.2ex}K}^{obs}$ and  $L_{C\MakeLowercase{a}\protect\scaleto{II}{1.2ex}K}^{*}$  are the observed and the expected luminosity of the \CaIIk\ line for a given {\halpha} luminosity, respectively, where the latter is given by eq.(\ref{halpha-cak}). This index reflects how far is a star from the general trend in Figure \ref{ctts-wtts}. The smaller the index, the smaller the observed luminosity is compared to the expected one, and thus the higher the depletion.

\begin{deluxetable}{lcc}
\tablecaption{Depletion index values for the selected sample of CTTS (see \S \ref{dep_index}) \label{tab:dep}}
\tabletypesize{\scriptsize}
\tablehead{\colhead{Object} & \colhead{i$_d$} & \colhead{$\sigma$ i$_d$}}
\startdata
CHX18N & -0.126 & 0.067 \\
CHXR 47 & 0.618 & 0.362 \\
CR Cha & -0.404 & 0.042 \\
CS Cha & -0.419 & 0.030 \\
CT Cha A & 1.019 & 0.152 \\
CW Cha & 0.051 & 0.057 \\
ESO-Ha-562 & 0.04 & 0.013 \\
J11085367-7521359 & -0.131 & 0.051 \\
J11432669-7804454 & -0.377 & 0.011 \\
Sz Cha & -0.291 & 0.096 \\
Sz18 & -0.278 & 0.017 \\
Sz19 & -0.238 & 0.115 \\
Sz22 & 0.222 & 0.051 \\
Sz27 & -0.811 & 0.007 \\
Sz32 & 0.334 & 0.052 \\
Sz33 & -0.144 & 0.030 \\
Sz37 & 0.41 & 0.052 \\
Sz45 & -0.445 & 0.020 \\
T10 & -0.589 & 0.010 \\
T12 & -0.344 & 0.018 \\
T16 & 0.778 & 0.232 \\
T23 & 0.154 & 0.014 \\
T24 & 0.092 & 0.048 \\
T27 & 0.058 & 0.038 \\
T28 & -0.494 & 0.010 \\
T3 & -0.129 & 0.038 \\
T3 B & 0.224 & 0.072 \\
T30 & 0.177 & 0.024 \\
T33 B & 0.01 & 0.102 \\
T38 & -0.11 & 0.035 \\
T4 & 0.275 & 0.044 \\
T40 & -0.011 & 0.063 \\
T44 & 0.38 & 0.043 \\
T45 & 0.134 & 0.037 \\
T46 & -0.775 & 0.010 \\
T48 & 0.114 & 0.012 \\
T49 & -0.617 & 0.053 \\
T52 & 0.093 & 0.028 \\
TW Cha & -0.048 & 0.025 \\
VW Cha & -0.604 & 0.007
\enddata
\end{deluxetable}

We report the $i_d$ values and their errors, calculated from observed and expected luminosity errors, in Table \ref{tab:dep}. Figure \ref{dep_hist} shows the histogram for the depletion indexes of the sample, the solid black line indicates the mean value (-0.05) obtained for the index, the black dashed lines show the values of the 16\% (-0.43), and 84\% (0.26) percentiles. We identify the stars below the 16\% percentile (i.e., stars with $i_d \leq -0.43$) as depleted; obtaining that 17.5\% of the selected sample is depleted. On the other hand, we note there is a \emph{tail} of stars in the distribution with $i_d$ greater than 2 times the 84\% percentile; however, these stars have high extinction values (Av $>$ 2) leading to high extinction-corrected fluxes at short wavelengths and in particular to cases with $L_{C\MakeLowercase{a}\protect\scaleto{II}{1.2ex}K}$ higher than $L_{H\alpha}$. Therefore, we require revisiting the extinction measurements of
CHXR 47, CT Cha A, Sz37, and T16. Future work with detailed profile modeling may clarify the nature of the rest of the stars with a high index value.

\subsection{Relationship to disk structure}

\begin{figure}[!t]
\epsscale{1.2}
\vspace{0.05in}
\plotone{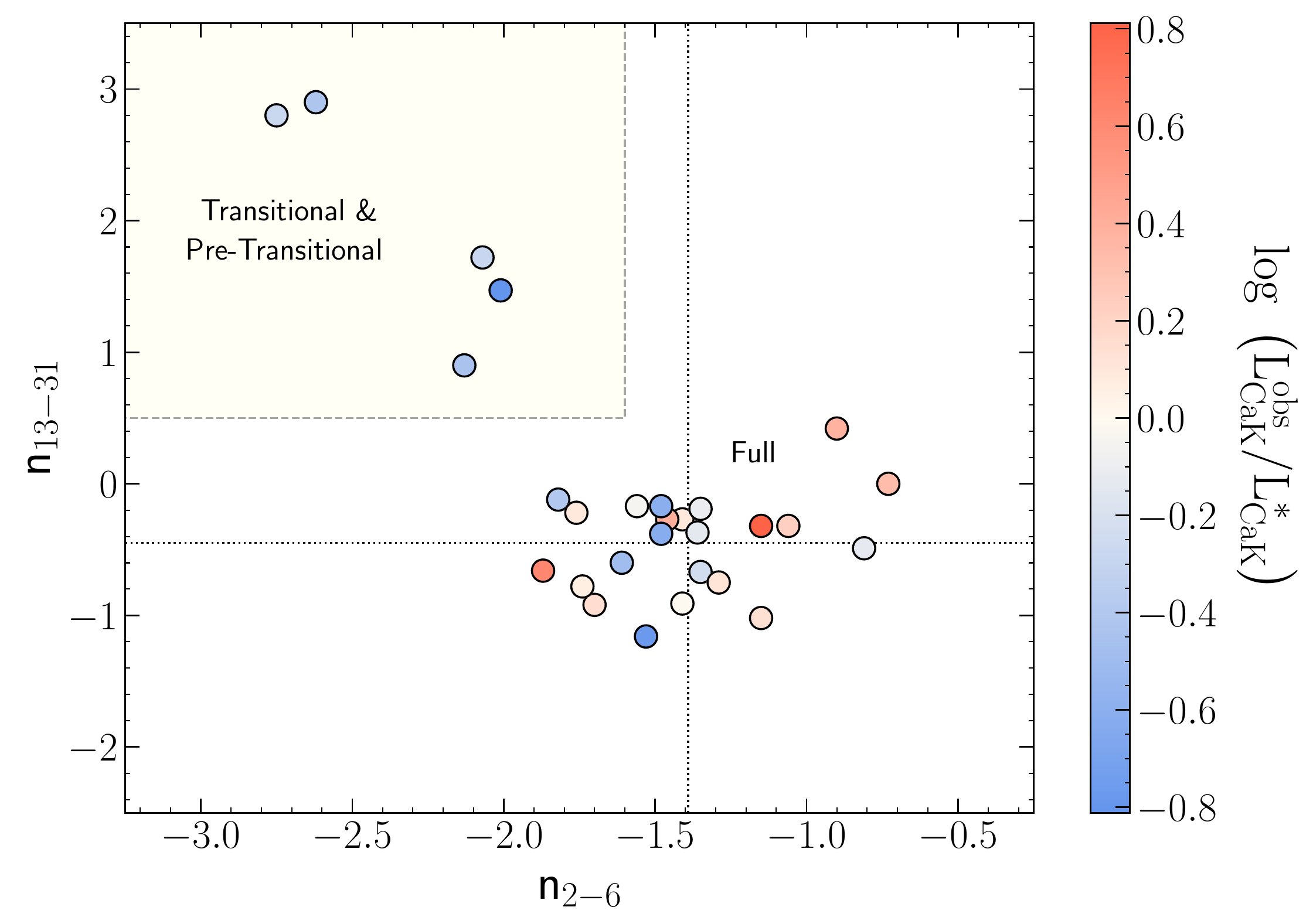}
\caption{Stars in Cha I in the continuum indices slopes n$13-31$ vs. n$2-6$ diagram. The light-yellow zone represents the location of the TD$/$PTD for the full Cha I sample in \citet[][]{manoj_spitzer_2011}. Error bars are smaller than the markers. The dotted lines represent the mean values for the full disks.  Colors indicate the degree of depletion of the star, the most depleted stars have low (blue) values of this index.
}
\label{slopes}
\end{figure}

We use the continuum spectral indices --- a measurement of the spectral energy distribution (SED) slopes --- between 2.159 $\mu$m and 5.7 $\mu$m  (n$_{2-6}$), and 13.4 $\mu$m and 31.1 $\mu$m (n$_{13-31}$), from \citet[][]{manoj_spitzer_2011} to explore the relationship between disk evolution and the depletion seen in the spectra.

In Figure \ref{slopes}, we show the location in the n$_{13-31}$ vs. n$_{2-6}$  diagram of 28 of our $L_{H\alpha}$ restricted sample, for which the spectral indices were available. We have indicated the distinct zones for full disks and transitional/pre-transitional disks, the yellow-colored region corresponds to the latter with boundaries given by \citet[][]{manoj_spitzer_2011}. A disk with no significant radial discontinuities in its dust distribution, a.k.a., full disk (FD), will have excess flux over the photosphere between $2-6\mu$m, mostly due to emission from the inner dust rim at the dust sublimation radius, or “dust wall'' \citep{dalessio_effects_2006}, which is directly illuminated by stellar radiation \citep{natta_reconsideration_2001, dullemond_passive_2001}. In contrast, the transitional disks \citep[TD, ][]{calvet_disks_2005} and pre-transitional disks \citep[PTD,][]{espaillat_confirmation_2008} have  disk cavities or gaps that decrease the near-IR emission relative to that from the optically thick FD. The cavities or gaps may or may not contain optically thin dust, and some optically thick material remains in the PTD, but overall the net effect is that the near-IR slope n$_{2-6}$ decreases relative to the FD. In addition, the edge of the cavity in TD/PTD is directly illuminated by stellar radiation resulting in higher emission in the mid-infrared than the observed for FD, increasing the n$_{13-31}$ slope relative to FD. These effects result in the displacement of the TD/PDT relative to the FD in the n$_{13-31}$ vs. n$_{2-6}$ diagram (see Figure \ref{slopes}).

We colored each star in Figure \ref{slopes} by their $i_d$ value, the smaller/bluer, the more depleted the CTTS. All TD/PTD show low $i_d$ values, suggesting there is a connection between disk evolution and the degree of depletion seen in the spectra. A number of FD also show low $i_d$ values; however, these disks also show hints of more evolved/processed dust, as they tend to concentrate below the n$_{2-6}$ mean, indicating decreased emission from the dust wall, which is the dominant contribution to the flux in the 2 to 6 $\mu$m range. In turn, this indicates that the dust in the wall is more settled towards the midplane \citep{manzo-martinez_evolution_2020}. We explore a possible relationship between n$_{2-6}$ and the depletion index for the full disks, obtaining a Pearson correlation coefficient of 0.30 suggesting a weak link may exist between the two. In addition, CR Cha, one of the high depletion, full-disk stars show enhanced silicate emission at 10 $\mu$m relative to the continuum emission, which has been interpreted as the stars being in the process of opening gaps, with the excess silicate emission coming from optically thin dust in the gap \citep[][]{manoj_spitzer_2011}.

\section{Modeling} \label{sec:models}

The depletion index defined in \S \ref{depletion} is a rough indicator of the degree of Ca depletion in the magnetospheric flow and thus in the inner disks of the stars. However, this does not provide an estimate of the Ca abundance in the flow relative to H. Modeling of the lines is needed for a precise estimate of the Ca abundance and therefore of the Ca depletion. In this section, we apply the magnetospheric accretion model to the Ca-poor star T28 (our benchmark for stars showing Ca depletion, c.f., Figure \ref{ca-poor profile}) to determine its Ca abundance.

\subsection{The magnetospheric accretion model}

We follow the model presented by \citet[][]{hartmann_magnetospheric_1994} and \citet{muzerolle_emission-line_1998, muzerolle_emission-line_2001}. In this framework, a dipole geometry is assumed for the magnetic field and the accretion flows, which follow the field lines, characterized by the inner radius (\ri) and the width at the base of the flow (\Dr) in the disk. The temperature is a free parameter, constrained by the accretion rate, following the prescription of \citet[][]{muzerolle_emission-line_2001}.

The models use the extended Sobolev approximation to calculate mean intensities, which in turn are used to calculate radiative rates in the statistical equilibrium equations. We consider a 16-level hydrogen atom \citep[][]{muzerolle_emission-line_2001} and a 5-level calcium atom. We assume solar abundances relative to H from \citet{asplund_cosmic_2005}, but introduce a new parameter $X$, which is the Ca abundance relative to the solar abundance. The line flux is determined by using a ray-by-ray method, in which the specific intensity emerging from the configuration and the total optical depth of each ray are calculated at a given inclination $i$. Line profiles are calculated using the Voigt function, with appropriate damping parameters. Finally, fluxes at each velocity along the line profile are calculated by integrating the specific intensity.

Based on the  behavior of the lines (\S \ref{sec: data}), we performed a simultaneous fit of multiple emission lines of hydrogen and calcium. With this approach, the H lines were used as the anchor to secure the \mdot\, while the fitting explores the Ca abundance needed to explain the disparity in the profiles. We calculated a large grid of profiles, having a total of 612080 profiles combined for all the lines, for a standard M1 star of 3 Myr, consistent with the estimated age for Cha I, and stellar properties derived from the PARSEC evolutionary models \citep[][]{bressan_parsec_2012}.
Table \ref{tab:model_param} describes the parameter space explored. Finally, to include the contribution of the chromosphere in our physical model, we add to the model line fluxes the corresponding line fluxes of a WTTS of the same spectral type as the CTTS being analyzed.

\begin{deluxetable}{lccc}[t]
\tablecaption{Range of Model Parameters \label{tab:model_param}}
\tablehead{
\colhead{Parameters} & \colhead{Min.} & \colhead{Max.} & \colhead{Step} 
}
\startdata
log {\mdot} ($\msunyr$)	& -10.0	& -7.0 	& 0.25  \\
T$_{\rm max}$ (K)	        & 6500	& 14000	& 500 \\
R$_{\rm i}$	(R$_{\star}$)	& 2.0	& 6.0	& 0.5 \\
$\Delta\rm r$ (R$_{\star}$)	& 0.5   & 2.0   & 0.5 \\
$i$	(deg)				        & 15	& 75	& 15 \\ 
X (solar)		& 0.001	&  1	& ... \\ 
\hline
\enddata
\end{deluxetable}

\subsection{Model fit to observed fluxes: the MCMC method} \label{sec: mcmc}

We used Bayesian inference to estimate the values of the magnetospheric parameters and Ca abundances, which produce the spectrum  which better fits the observed line fluxes. In addition to the line fluxes, we use the FWHM of the \halpha\ line ($W_{{\rm H}\alpha}$) as an extra data point to fit, which allows us to include information about the profile, important to constrain the inclination of the line of sight and the geometry of the magnetosphere.
We use a likelihood given by:

\begin{equation}
\begin{split}
    & \log\ \mathcal{L} \propto  \\ & - \frac{1}{2} \left( \frac{{(W_{{\rm H}\alpha,obs} - W_{{\rm H}\alpha,mod})}^2}{\sigma_{W_{{\rm H}\alpha}}^2} + \sum_{i} \frac{(F_{i,obs} - F_{i,mod})^2}{\sigma_{i}^2} \right) 
\end{split}
\end{equation}

\noindent 
Where $F_{i,{\rm obs}}$,  $F_{i,{\rm mod}}$ are the observed and the model fluxes, respectively, and $\sigma_i$ is the flux error. The subscript $i=1,...,6$ refers to the three Balmer (\halpha, \hbeta, \hgamma) and the three \CaII\ lines ($\rm K$, $~\lambda$8498, $~\lambda$8542).

To sample the posterior probability density function (PPDF), we use the \emph{emcee} implementation module \citep[][]{foreman-mackey_emcee_2013}; the computation ran for 100 walkers with 3000 steps each one, from which we expect convergence for all parameters. To speed up the calculation process, from the models' grid we assemble a 6D piecewise interpolants to obtain the total fluxes of the models corresponding to each line. In each realization, the MCMC was left to explore freely the parameter space of the models with the only exception of the variable log \mdot, where we used a Gaussian distribution of $\sigma = 0.1$, centered in the value of the mass accretion rate obtained in M17.

We use the median of the marginal posterior PDFs to estimate the model parameters  which provide the best fit to the observed fluxes. The  16\% and 84\% percentiles are taken as the limits of the credibility interval.

\subsection{Model Results for T28}

\begin{deluxetable}{lcc}[!t]
\tablecaption{Results of Model Parameters for T28 \label{tab:model_results}}
\tablehead{
\colhead{Parameters} & \colhead{Inferred value}
}
\startdata
\vspace{1mm}
log {\mdot} ($\msunyr$)	& -8.14 $^{+0.21}_{-0.22}$  \\
\vspace{1mm}
T$_{\rm max}$ (K)	 & 8231.51 $^{+371.44}_{-256.91}$ \\
\vspace{1mm}
R$_{\rm i}$	(R$_{\star}$) & 3.15 $^{+0.33}_{-0.23}$  \\
\vspace{1mm}
$\Delta\rm r$ (R$_{\star}$)	& 0.74 $^{+0.13}_{-0.11}$ \\
\vspace{1mm}
$\cos{i}$			        & 0.67 $^{+0.03}_{-0.07}$ \\
\vspace{1mm}
X (solar)		& 0.17 $^{+0.10}_{-0.06}$	 \\ 
\hline
\enddata
\end{deluxetable}

\begin{figure*}[!t]
\epsscale{0.92}
\vspace{0.05in}
\plotone{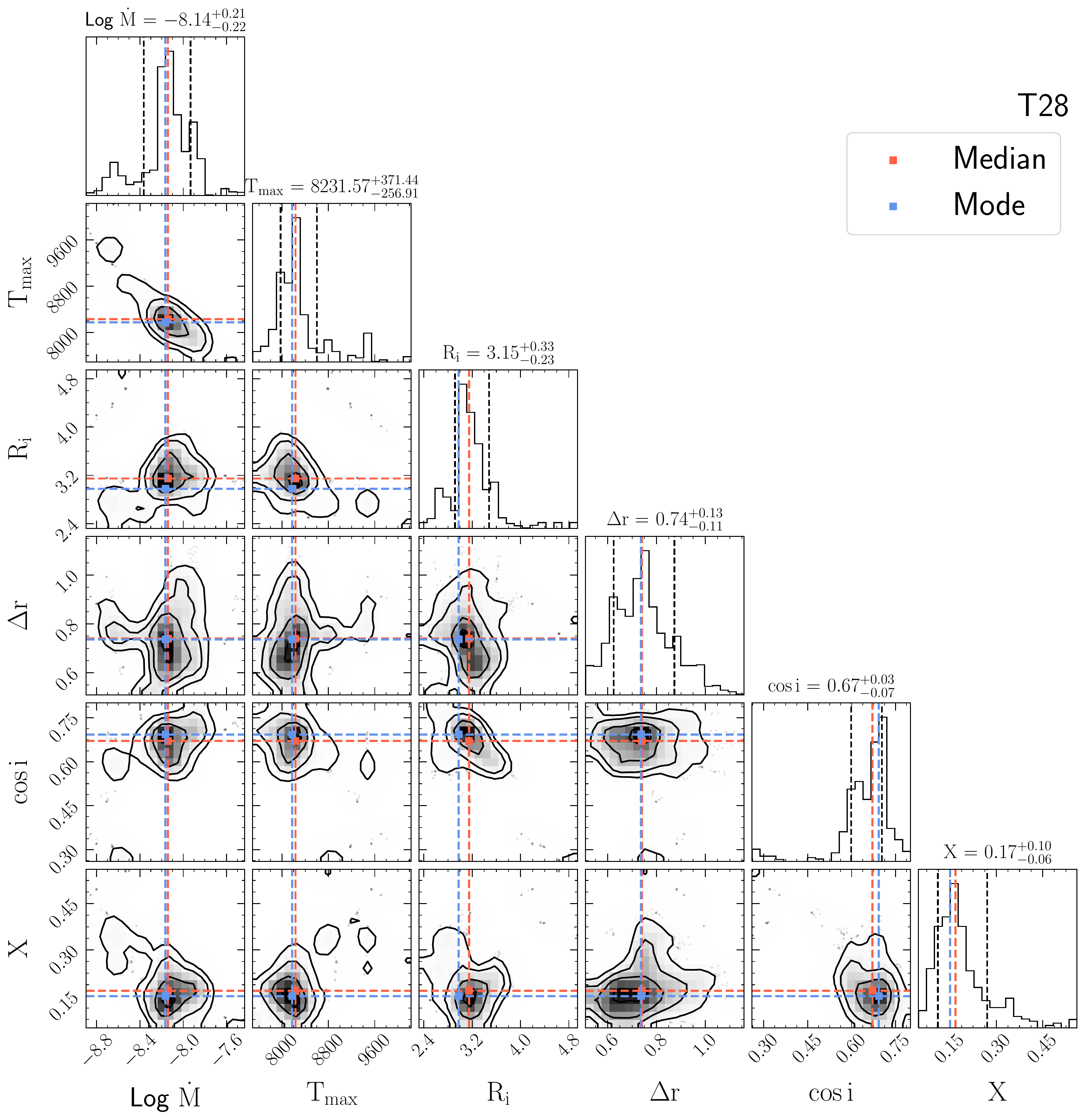}
\caption{MCMC corner plots tracing off magnetospheric accretion parameters and Ca abundance for T28. Red and blue dashed lines represent the median and mode values of the distribution, respectively.  The values reported in the figure correspond to the median and the credible intervals defined from 16th and 84th quartiles (black dashed lines), see Table \ref{tab:model_results}.}
\label{corner}
\vspace{0.016in}
\end{figure*}

\begin{figure*}[!t]
\vspace{0.05in}
\centering
\epsscale{0.95}
\plotone{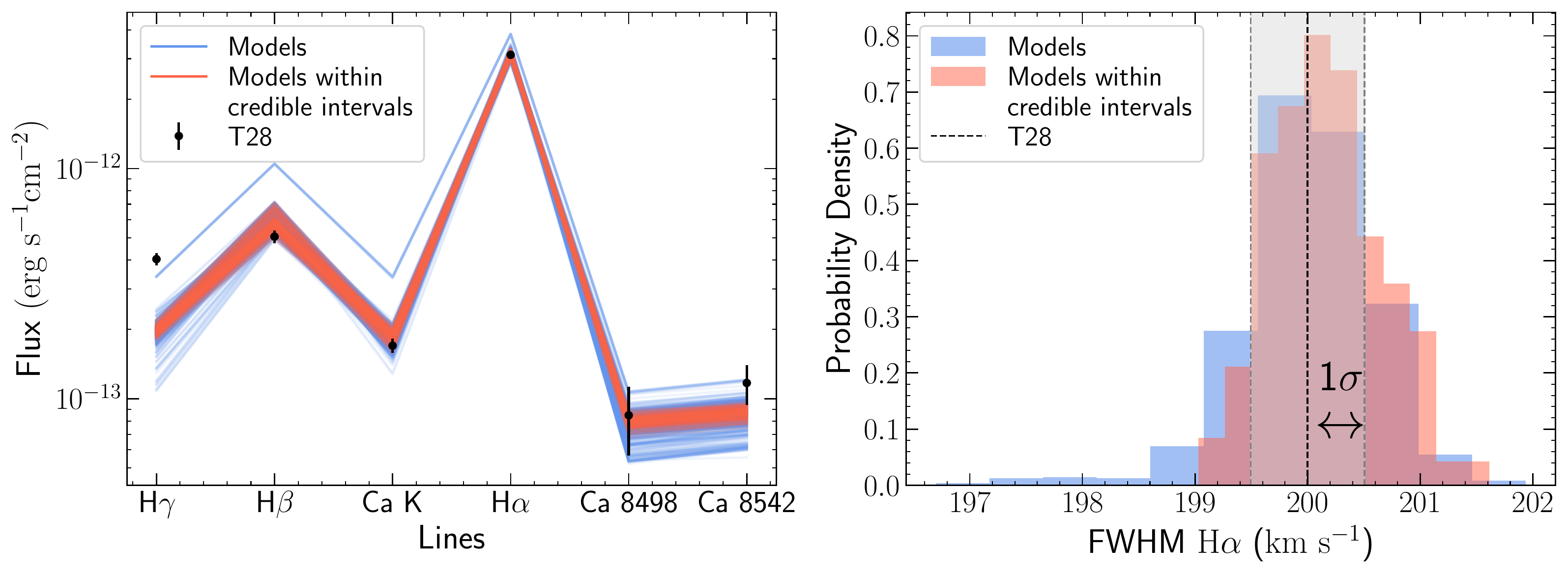}
\caption{\textit{Left:} 
Fit of 1000 random models (see text) to the fluxes of \halpha, \hbeta, \hgamma, \CaIIk, \CaIItwfi, \CaIItrfi\ for T28 (blue lines). We highlight in red the models that fall within the credible intervals. \textit{Right:} Probability density distribution of the FWHM for the \halpha\ line for the models showed in left panel.}
\label{fits}
\end{figure*}

We apply our MCMC method to T28, using the WTTS TWA13B as the proxy for the chromosphere contribution to the lines. The results are shown in Figure \ref{corner}, showing the posterior probability distribution of log \mdot, \tmax, \ri, \Dr, \cosi\ and \rm{X}.  Each parameter is  well-constrained with sharply peaked marginalized distributions. Table \ref{tab:model_results} shows the median values for each parameter and their credible intervals. The fitting details are shown in Figure \ref{fits}. The left panel shows in blue the fit to the line fluxes of the 1000 random models and in red the fit to the subset of models that fall within the credible intervals defined from 16th and 84th quartiles. The right panel shows the probability distribution of the FWHM of \halpha\ for the same groups of models.

We note that the models are unable to successfully reproduce the fluxes for \hgamma. This discrepancy between  models and  observations, rather than being a consequence of the method used, seems to be a systematic limitation of the models; in fact, reproducing the Balmer decrement has proved problematic in the past even with the improvement made in the models \citep[e.g.][]{muzerolle_emission-line_2001}. As a possible explanation for the discrepancies seen for \hgamma, we explored the role of the temperature structure distribution in our calculations, as it remains one of the major uncertainties of the models \citep[][]{muzerolle_emission-line_2001}. To do so, we calculated the Balmer decrement for different spatial distributions of the heating rate \citep{hartmann_magnetospheric_1994}, and found no significant changes in line fluxes, because the hydrogen lines form in 
similar regions. The discrepancies between the models and observations may be due to intrinsic departures from the axisymmetric geometry assumed in our models, which requires a more detailed analysis to disentangle; this however is beyond the scope of this paper. Nevertheless, we are able to reproduce simultaneously \halpha\ and \hbeta\ with the calcium lines, and we obtain a 0.17 solar Ca abundance in the magnetospheric flows with a mass accretion rate of log $\Mdot = -8.14$, consistent with the UV-determined M17 value, within the uncertainties.

\section{Discussion} \label{sec:discussion}

We have presented an observational study of the calcium II lines in CTTS in the Cha I star-forming region. The samples presented here include objects at different disk evolutionary stages, i.e., objects surrounded by full disks, and by transitional disks, and covering a fairly wide range of accretion rates, making it a reasonably complete sample of young stellar objects. We find a range of Ca depletion in the sample, with significant depletion in disks in advanced stage of evolution, possibly including planet formation. In the following, we discuss some aspects related to our results.

\subsection{The chromospheric component of the Ca lines}

The \CaII\ lines are important secondary tracers of accretion as the strong trends between the line luminosities and the accretion rate show, also found in previous work \citep[e.g.][]{ingleby_accretion_2013}. However, the results presented in \S3 show that the lines show an significant chromospheric contribution that becomes the main component of the lines as \mdot\ decreases. In the low \mdot\ cases, the lines still show traits of magnetospheric accretion, such as red-shifted absorption and a ghost of the high velocity wings, the latter mostly in the \CaIIk\ line. In any case, the chromospheric component only becomes negligible at $\Mdot \geq -8.3\ \msunyr$, 
but then again this limit must be taken carefully and an inspection of the profiles prior to any analysis is recommended even for high accretors.

A factor to consider is how relevant is to include the chromosphere is the newly-found depletion of Ca in the stellar spectra. Although relatively easy to spot in high accretors (such as T28), depletion can also be present in stars with low \mdot, but it is more difficult to identify given that their profiles are naturally more chromospheric. Ideally, the depletion index we defined can get around this problem; however, the results may be erroneous because of uncertainties in extinction or having lower apparent fluxes because of strong red-shifted absorption in the profiles. Nevertheless, an inspection of the line profiles can solve most problematic cases.

Overall, the  disks identified as depleted by their profiles match the ones identified by the depletion index. Despite its approximate nature, it allows us to reasonably make a coarse examination of the depletion status of a stellar sample. However, detailed modeling is needed to actually measure abundances.

\subsection{Reference abundance}\label{sec:reference}

In the abundance determination of T28 using magnetospheric accretion modeling (\S \ref{sec:models}) we use as reference the currently most accepted solar abundances \citep{asplund_cosmic_2005}, which differ from the recommended solar abundances \citep{lodders_solar_2010} only in the volatiles, especially oxygen. The reference abundance should be that of the original cloud, before it got changed by processes happening in the disk. That abundance is not directly available, but we infer it by making two assumptions: first, that the stellar abundances represent those of the cloud, and second, that the stellar abundances are solar.

That the abundances in the stellar surface are representative of the cloud abundances is justified because the Cha I stars we study are low mass stars on the Hayashi track of pre-main sequence evolution and therefore are fully
convective.  Convection rapidly mixes freshly accreted material with the bulk of stellar material \citep{jermyn_stellar_2018,kunitomo_revisiting_2018} and washes out any abundance anomalies, so that the stellar abundances are representative of the earliest abundances in the disk, when it formed along with the star. This situation is different from the young Herbig Ae stars, in which the lack of convection and mixing implies that stellar abundances retain the imprint of accretion \citep{cowley_abundances_2010,cowley_chemical_2014}.

We also assume that young stars have solar
abundance.  The current solar abundances differ from the original abundances in the protosun because of the gravitational settling of He and heavy elements
\citep{lodders_solar_2003}; however,  \citet{lodders_solar_2010} finds differences of $<$ 0.1 dex between the Ca  abundances in the protosun and the current solar abundances. Direct determination of abundances in young stars does not depend on models for extrapolating back in time, but they are difficult for low-mass young stars because of several effects such as veiling of the spectral lines due to accretion shock emission, magnetic effects, and to lesser extent rotation. 
Nonetheless, abundances have been obtained for stars nearby star-forming regions. In particular, the metalicity of the  Chamaeleon I star-forming region has been found slightly subsolar but still compatible with the solar values within the uncertainties (Santos et al 2008, Spina et al. 2014). This chemical pattern was also found in the young stars of other star-forming regions like the Taurus molecular cloud \citep{padgett_atmospheric_1996,santos_chemical_2008,dorazi_chemical_2011}, and the Orion complex \citep{santos_chemical_2008,dorazi_metallicity_2009,biazzo_chemical_2011,biazzo_elemental_2011}. Since Taurus and Orion represent two different modes of star formation, in that Taurus includes distributed populations of low-mass stars while Orion includes clustered populations with high-mass star formation, we assume that solar chemical abundances are representative of the solar neighborhood.

Given these considerations, the chemical differences we found between the stars cannot be explained by the stars having different parent clouds with distinct chemical compositions. One possibility to explain the differences could be chromospheric activity, since metal abundances in stellar spectra seem to correlate with magnetic activity, especially in young active stars like TTS \citep[e.g.][]{flores_discovery_2016,galarza_effect_2019,spina_how_2020}. However, for the calcium lines the effects of stellar activity due to equivalent width and atmospheric microturbulence cancel out each other, which results in a negligible change in their abundances \citep[][]{spina_how_2020} and suggests that the chemical anomalies we found are genuine.

\subsection{Nature of Ca depletion}

According to the considerations of the previous section, the observed depletion most likely originated in the disk, before the material reached the magnetospheric flows. In a full disk, the large grains --- a.k.a. “pebbles'' --- are settled in the midplane and will spiral inwards onto the star. As they drift radially inward, the pebbles encounter higher and higher temperatures, leading elements to sublimate and return to the gas. However, refractory elements need temperatures of $\sim$ 1500 K to sublimate, so they will be in the solid phase until they reach those high temperatures at the dust wall at the edge of the dust disk; hence, refractory elements are not affected by typical loss mechanisms such as MHD winds and/or photo-evaporation, and these mechanisms cannot be the reason for the abundance differences found in our analysis.

For calcium (or any other refractory element) to be depleted in the gas, it must be trapped outside its sublimation radius, i.e., outside the dust wall. Trapping of large dust grains induced by radial gaps and cavities in disks has been shown to correlate with  depletion of refractory elements like iron in Herbig Ae stars \citep[][]{kama_fingerprints_2015}. The cavities, i.e.,  regions cleared of small dust in the transitional disks, were the first structures identified in protoplanetary disks \citep[][]{espaillat_observational_2014}, and were attributed to dust trapping in the outer disk by giant planets
\citep[][]{whipple_certain_1972,barge_did_1995,rice_planetesimal_2006,pinilla_trapping_2012, zhu_dust_2012}. We now know that structures are probably present in most disks, and ringed structures in particular are ubiquitous in disks around stars of all spectral types \citep[][]{andrews_disk_2018,andrews_observations_2020}, and although planet formation is still a favored mechanism to explain the structures in disks, other effects may also be at play, such as photoevaporation
\citep[][]{alexander_photoevaporation_2006-1,alexander_photoevaporation_2006, owen_protoplanetary_2011,owen_theory_2012} accumulation of gas and dust at the outer edge of a region with low ionization (“dead zones'') \citep[][]{regaly_possible_2012, flock_gaps_2015}, and condensation fronts of different molecular species \citep[][]{zhang_evidence_2015}. 

Either way, the mechanism by which the  refractory abundance is lowered in the inner disk in disks with cavities is similar: a pressure bump is formed in the disk, and larger dust particles accumulate in the local pressure maximum \citep[][]{pinilla_trapping_2012,birnstiel_dust_2016}, creating the substructures we commonly see in young stellar objects.

As large grains dominate the dust mass, trapping has a major effect on the total elemental composition of the disk material that reaches the star, affecting not only the refractory material \citep[][]{kama_fingerprints_2015} but also freeze-out volatiles, as long the trapping occurs further out their corresponding snow-line \citep[][]{mcclure_carbon_2019}. Recent modeling work in dust/disk evolution including chemistry supports this hypothesis, showing planet-originated dust-traps are very efficient at blocking the pebbles exterior to their orbits, preventing them from reaching the host star and be accreted. Moreover, planets build their cores from planetary seeds formed by accreting pebbles while migrating inwards through the disk, until they reach the mass needed to accrete gas \citep[e.g.][]{johansen_forming_2017, ndugu_probing_2021, schneider_how_2021-1, schneider_how_2021}. Additionally, their atmospheres can also be enriched via collisions  or via the accretion of planetesimals during both phases. Therefore, a large refractory content in planetary atmospheres might be a sign of additional solid pollution, i.e., the addition of solids into the atmosphere via planetesimals, giant impacts, or dust transported through meridional flows during the gas-disk phase \citep[e.g.][]{ogihara_formation_2021, schneider_how_2021, lothringer_new_2021}, and has been suggested as a possible explanation for the refractory content in the planetary atmosphere of Jupiter and Saturn \citep[][]{schneider_how_2021}

Although the extent of the effects depends on the disk’s viscosity and location of the planet(s), it is clear that pressure bumps and planet formation may be associated with changes in the inner disk chemistry affecting the composition of the material falling onto the star. This view is reinforced by the fact all the disks with large gaps, i.e., the transitional/pre-transitional disks, in our sample show depletion.

The most peculiar objects of our sample are those showing depletion and no indication of planet formation in their disks. In particular, our prototype object T28 belongs to this category. Some, but not all, of these objects show some degree of dust-processing, hinting that they may be in the way to developing  a gap yet still not noticeable on the SED. High angular resolution imaging in the sub-mm has been exceptional in discerning substructures in disks; however, there is no high-resolution imaging for Cha I so far, and we cannot test this hypothesis. Nonetheless, \citet[][]{schneider_how_2021-1} find that the heavy element abundance in the gas that reaches the inner disk can be lower than solar by factors of 10 to 1 \% in the first few years of the disk, even in disks that have not formed planets, consistent with the depletion we find for Ca in T28. This is due to the fact that with time, the heavy element content in the inner disk decreases because the gas in the innermost disk, initially enriched by sublimation of  the material transported inwards by pebbles, is accreted onto the star and the gas that arrives later is depleted of heavy elements \citep{schneider_how_2021-1}. These processes are very dependent on the assumed viscosity parameter. For instance, at low viscosities, pebbles grow larger and drift inwards faster \citep{birnstiel_simple_2012}, so the inner gas disk gets enriched of elements in the solids and is accreted onto the star sooner than in high viscosity disks \citep{schneider_how_2021-1}. Models of disk evolution including predictions on how the inner disk abundances of different elements change with time for ranges of characteristic parameters  are needed in order to compare with our determinations.

Another possible explanation for refractory depletion is the formation of Systems with Tighly-packed Inner Planets (STIPs) also called Earths or super-Earths, given their mass. The orbital radii of these planets are inside $\sim$ 1 au \citep[][]{tan_overview_2016} and moreover, a significant fraction of these planets are inside $\sim$ 0.1 au, which locates them directly in the inner gas disk. One of the possible explanations for STIPs is that they form in-situ in the gas \citep[][]{chiang_minimum-mass_2013, hansen_migration_2012, chatterjee_inside-out_2014}, if this is the case, most of the refractory materials --- which form the bulk of the rocky material in earth-like planets \citep[][]{lodders_mike_2003} --- including Ca, must go to the planets and gets depleted in the gas and consequently, in the magnetospheric flows.

Future work is needed to discern the nature of the depletion, including refractory abundance determinations for large samples of stars, combined with modeling of SEDs and dust spatial distributions, to characterize the observed depletion in the innermost disks. These will put strong constraints on models of disk and dust evolution; in addition, variation in elemental abundances can potentially give insight into both the composition of planets and their formation.\\

\section{Summary and conclusions} \label{sec:conclusions}

We used VLT X-shooter and Spitzer IRS observations to analyze the Ca II lines in a large sample of Chameleon I T Tauri stars. The main results are summarized here.

\begin{enumerate}

    \item The {\CaII} lines have an important  chromospheric component that increases in relevance as the mass accretion rate decreases, becoming equally important than the magnetospheric counterpart for $\log\ \left( {L_{Ca II K}}/{\lsun} \right) \leq -3.4$. 
    
    \item We found evidence of Ca depletion in the magnetospheric accretion flows (and therefore in the inner disks) of some stars in the sample. This is detected by profile inspection for high accretors and roughly measured by an index based on line fluxes; however, low accretors require a detailed analysis of line profiles.
    
    \item All disks with strong indications of dust evolution --- the transitional and pre-transitional disks --- in our sample exhibit Ca depletion, establishing a clear connection between inner disk refractory composition and outer disk structure in T Tauri stars. We found that a fraction of full disks in our sample show depletion in their spectra and also hints of more processed dust; more detailed analysis is required in these cases.
    
    \item As a prototype analysis, we performed a simultaneous fitting of the H and \CaII\ lines for the Ca-poor star T28, obtaining a calcium abundance of about 0.17 the solar abundance for the magnetospheric flows/inner disk. Such depletion cannot be explained by 
    differences in initial cloud abundances or chromospheric activity, suggesting this effect must originate in the disk. Future work will implement a similar analysis for the rest of the sample.

\end{enumerate}

\acknowledgments
\noindent
We thank Carlo Manara for providing the X-shooter data for the Cha I sample. 
This research was partially supported by NASA grant NNX17AE57G and HST grant AR-16129.\\
\software{Astropy \citep[][]{astropy_collaboration_astropy_2013,astropy_collaboration_astropy_2018}, PyAstronomy \citep[][]{czesla_pya_2019},  Eniric \citep[][]{neal_jason-nealeniric_2019}, emcee \citep[][]{foreman-mackey_emcee_2013}, scipy \citep[][]{virtanen_scipy_2020}.}

\bibliography{micolta2022.bib}{}
\bibliographystyle{aasjournal}

\appendix
\restartappendixnumbering
\section{Ca-poor stars}

Here, we present the profiles for the \halpha\ \CaIIk and \CaIItrfi lines of the stars identified as Ca-poor, as discussed in section \ref{chro}.
In all figures, the stars are sorted by the value accretion rate from M17, the profiles are shown prior to the subtraction of the photosphere.

\begin{figure*}[h!]
\epsscale{0.8}
\vspace{0.2in}
\plotone{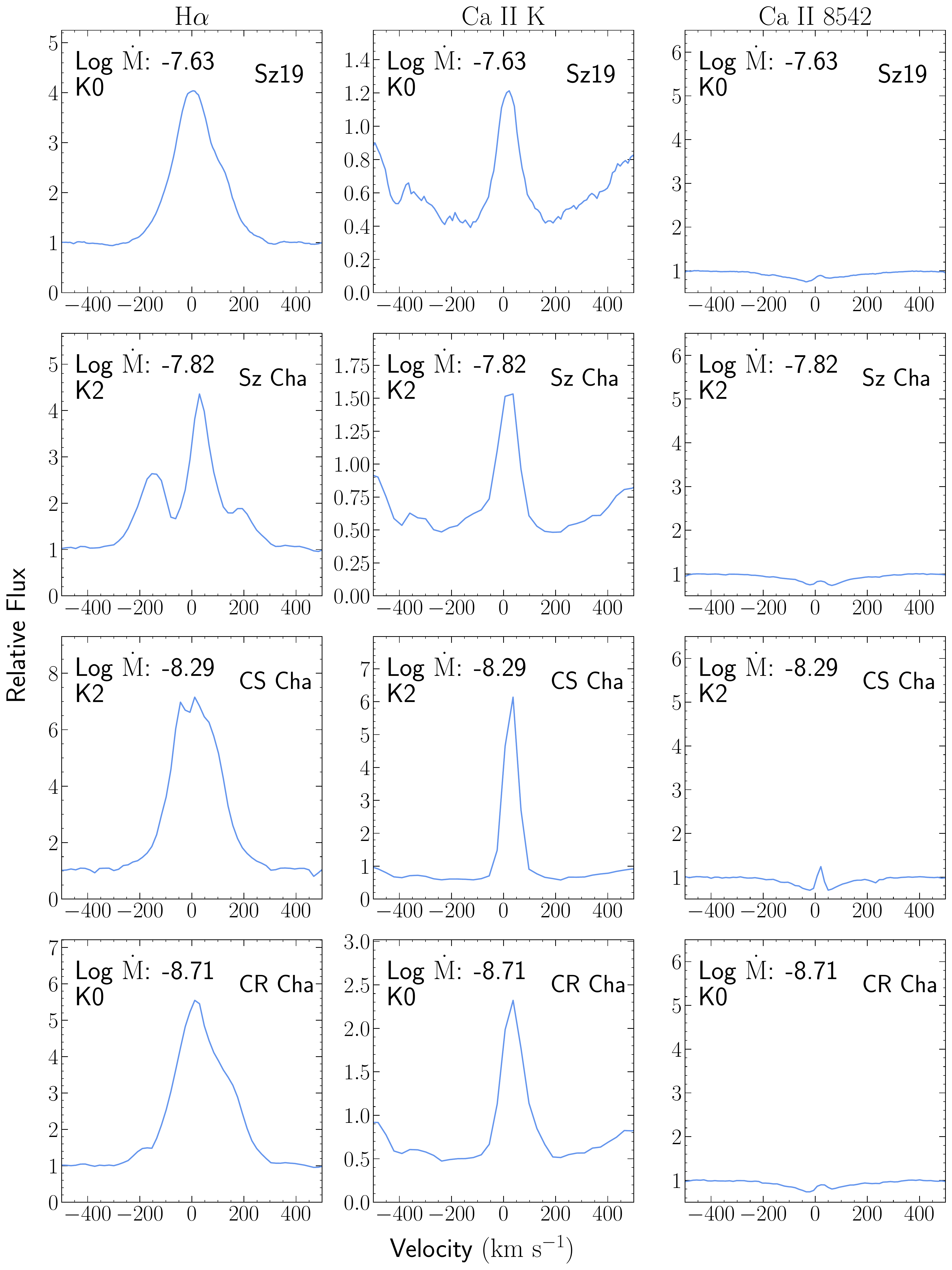}
\caption{ Profiles of \halpha, \CaIIk, and \CaIItrfi, one of the IR triplet lines, for the stars earlier than K7 identified as Ca-poor.}
\label{fig-a: prof 1}
\end{figure*}

\begin{figure*}[h!]
\epsscale{0.8}
\vspace{0.2in}
\plotone{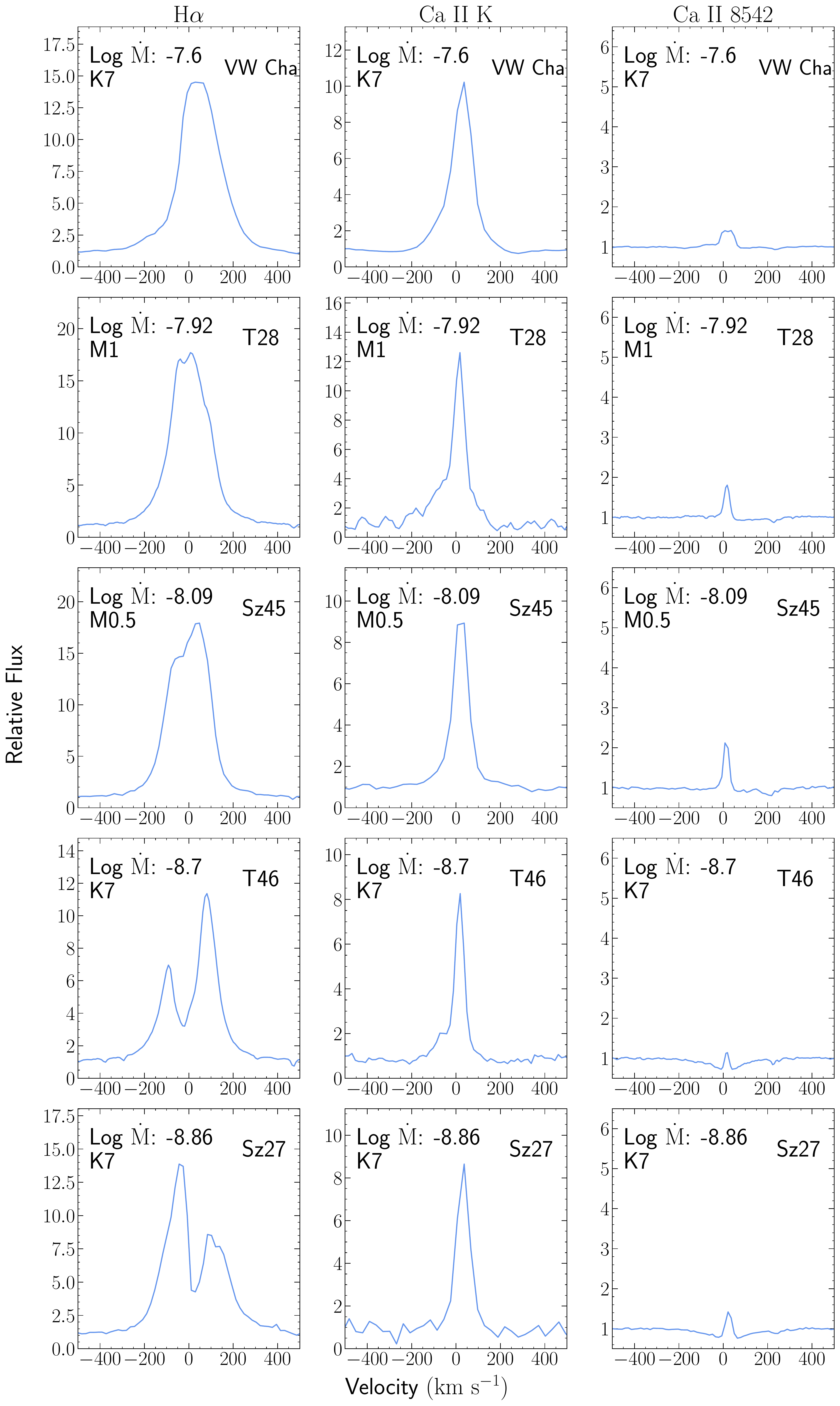}
\caption{Profiles of \halpha, \CaIIk, and \CaIItrfi, one of the IR triplet lines,  for stars later than K7 but  equal or earlier than M1 identified as Ca-poor.}
\label{fig-a: prof 2}
\end{figure*}

\begin{figure*}[h!]
\epsscale{0.8}
\plotone{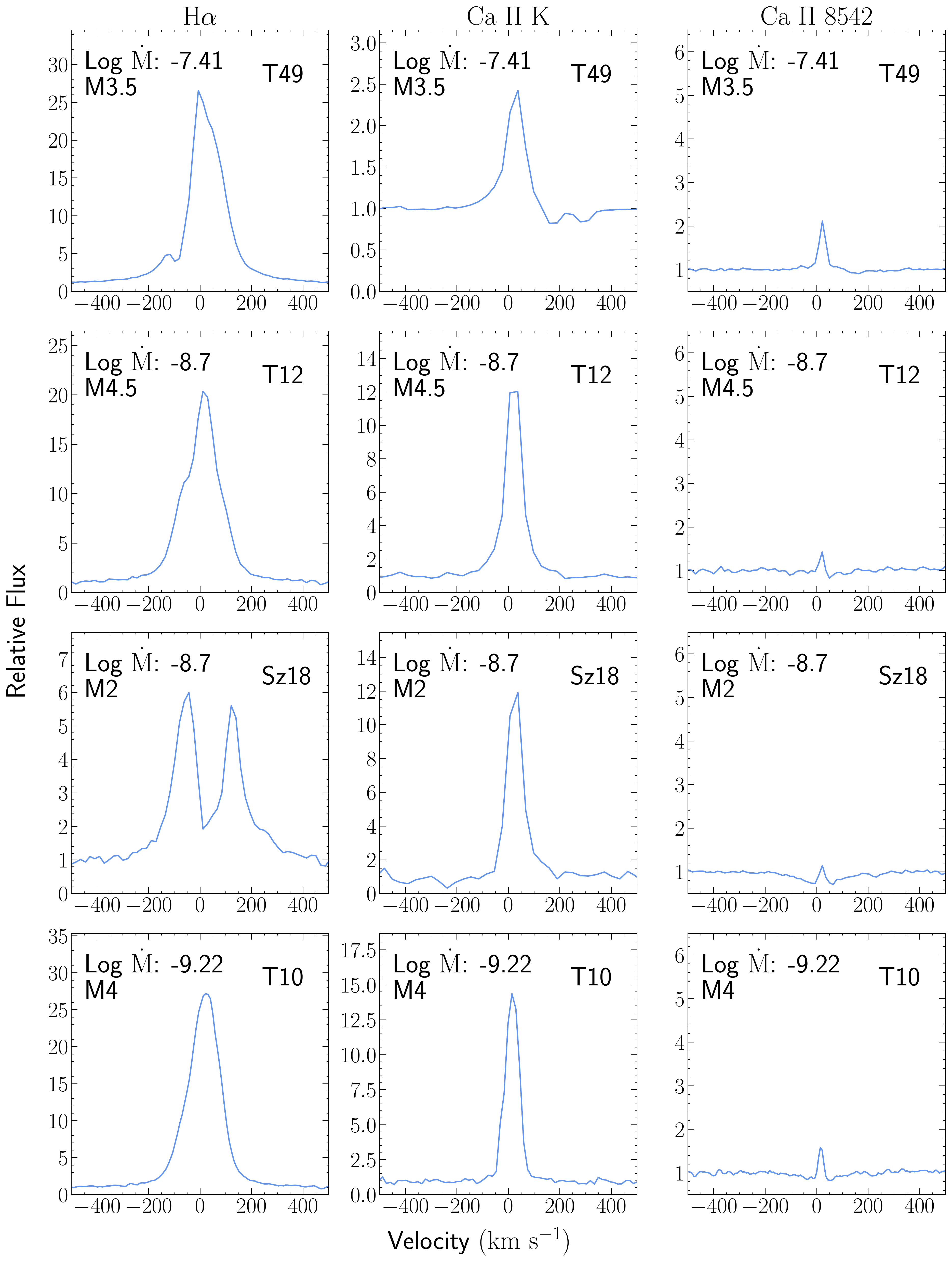}
\caption{Profiles of \halpha, \CaIIk, and \CaIItrfi, one of the IR triplet lines, for stars later than M1 identified as Ca-poor.}
\label{fig-a: prof 3}
\end{figure*}

\begin{figure*}[h!]
\epsscale{0.8}
\plotone{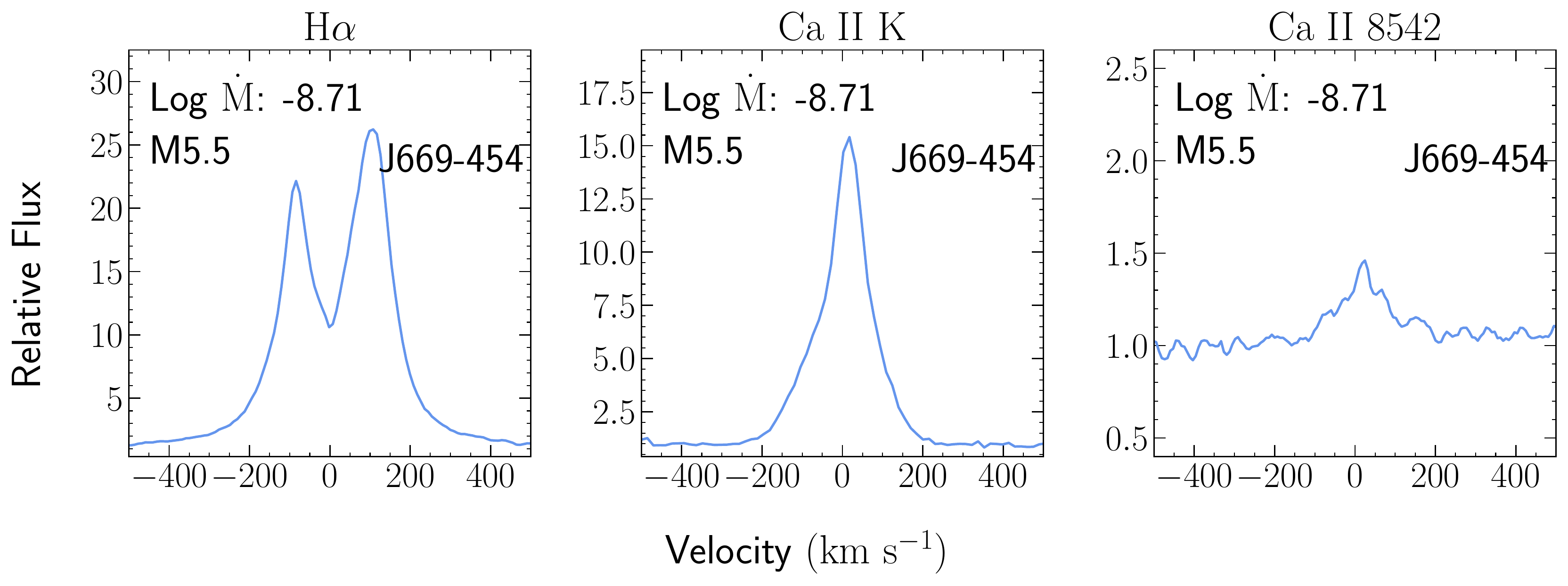}
\caption{Profiles of \halpha, \CaIIk, and \CaIItrfi, one of the IR triplet lines, for J11432669-7804454.}
\label{fig-a: prof 4}
\end{figure*}

\end{document}